\documentclass[12pt]{article}
\usepackage{amssymb}
\usepackage{graphicx}
\usepackage{amsmath}
\usepackage[counterclockwise]{rotating}
\usepackage[hidelinks]{hyperref}
\hypersetup{colorlinks=true,citecolor=blue,pdfborder=000, urlcolor=red}
\usepackage{graphicx} 
\usepackage{threeparttable}
\usepackage[round,semicolon]{natbib}  
\usepackage{cite}  
\usepackage{longtable}
\usepackage{rotating}
\usepackage{fancyhdr} 
\usepackage{epstopdf}
\usepackage{tikz}
\usepackage{caption}
\usepackage{subcaption}
\usepackage{natbib} 
\usepackage{booktabs} 
\usepackage{array} 
\usepackage{verbatim} 
\usepackage{fancyhdr} 
\usepackage{float} 
\usepackage{adjustbox}
\usepackage{pdflscape}

\usepackage{indentfirst}
\usepackage{setspace,caption}
\usepackage[tmargin=1in,bmargin=1in,lmargin=1in,rmargin=1in]{geometry}
\usepackage{titlesec}
\titleformat*{\section}{\fontsize{16}{18}\bfseries\sffamily}
\titleformat*{\subsection}{\fontsize{14}{16}\bfseries\sffamily}
\titleformat*{\subsubsection}{\fontsize{12}{14}\bfseries\sffamily}

\usepackage{tabularx}

\begin{document}
	
\title{\Large Liquidity Constraints, Cash Windfalls, and Entrepreneurship: Evidence from Administrative Data on Lottery Winners}

\author{\normalsize Hsuan-Hua Huang\thanks{Department of Economics, Washington University in St Louis; Email: h.hsuan-hua@wustl.edu}, Hsing-Wen Han\thanks{Department of Accounting, Tamkang University; Email: hwhan466@mail.tku.edu.tw}, Kuang-Ta Lo\thanks{Department of Public Finance, National Chengchi University; Email: vancelo@nccu.edu.tw},  Tzu-Ting Yang\thanks{Institute of Economics, Academia Sinica, Email: ttyang@econ.sinica.edu.tw.} \medskip}

\maketitle

\begin{abstract}
	
	Using administrative data on Taiwanese lottery winners, this paper examines the effects of cash windfalls on entrepreneurship. We compare the start-up decisions of households winning more than 1.5 million NT\$ (50,000 US\$) in the lottery in a particular year with those of households winning less than 15,000 NT\$ (500 US\$). Our results suggest that a substantial windfall increases the likelihood of starting a business by 1.5 percentage points (125\% from the baseline mean). Startup wealth elasticity is 0.25 to 0.36. Moreover, households who tend to be liquidity-constrained drive the windfall-induced entrepreneurial response. Finally, we examine how households with a business react to a cash windfall and find that serial entrepreneurs are more likely to start a new business but do not change their decision to continue the current business.

\end{abstract}

\enlargethispage{2\baselineskip}%
\thispagestyle{empty}\newpage \setcounter{page}{1}\baselineskip=24pt

\section{Introduction}
New firms are usually considered as the drivers of economic growth and job creation \citep{ayyagari, haltwanger, kirzner, adelino}, and so promoting entrepreneurship is one of the key issues for many countries across the world. In order to encourage entrepreneurial activities, governments implement a variety of policy tools, such as providing seed money to start-ups, subsidizing loan payments to small businesses, or reducing the corporate tax burden. The rationale behind these policies is that limited credit availability and sizable capital requirements constrain low-wealth households from starting their own business. 

\begin{figure}[h!]
	\caption{Business Entry Rate and Household Wealth Percentile}\label{wealth_b}
  	 \centering
   	 \includegraphics[width=0.9\textwidth]{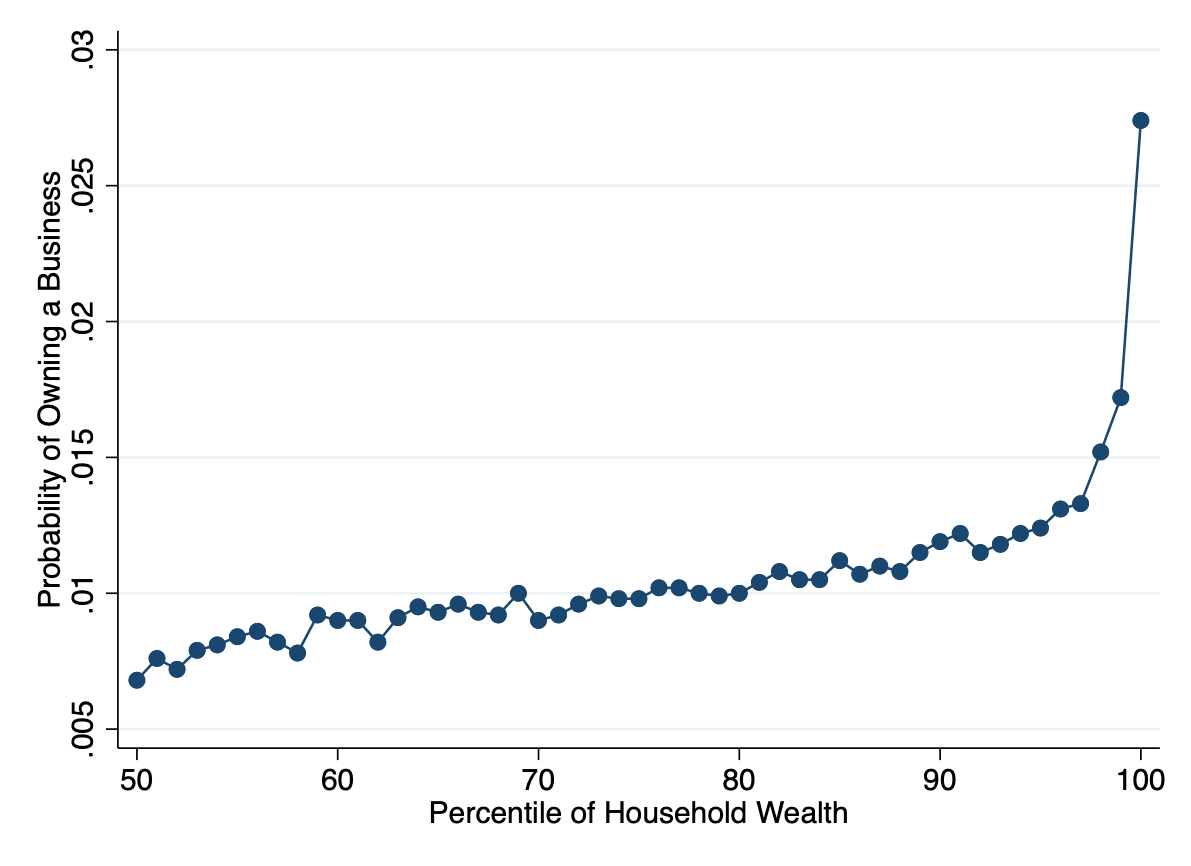}
	 \vspace{0.3cm}
	\vspace{0.3cm}
	\begin{minipage}{1\linewidth}		
	\fontsize{10}{10pt}\selectfont
	\emph{Notes:} 
	The figure displays the relationship between probability of owning a business and percentile of household wealth. Sample includes all households in Taiwan that did not own a business in the previous year and whose members are over 20. The horizontal axis is percentile of total household wealth. The vertical axis is probability of owning a business.
	\end{minipage}
\end{figure}

The seminal paper by \citet{evansdavid} argued that if a capital market is experiencing friction, a lack of financial resources (i.e., wealth) could prevent a skilled entrepreneur from starting a new business, thus implying that entrepreneurial entry
should respond positively to an increase in household wealth \citep{evansdavid, evans, fairlie1999absence, quadrini1999importance, glenn2004entrepreneurship}. Figure \ref{wealth_b}, using our administrative data, presents cross-sectional evidence on the business ownership and household wealth percentile.\footnote{Sample includes all households in Taiwan that did not own a business the previous year and whose members are over 20.} We find that the probability of owning a business increases in line with household wealth level, albeit this positive relationship might not represent the causal effect of wealth on business formation, since it is also possible that being an entrepreneur will help a household accumulate more wealth, or there is a positive selection due to other confounding factors such as family background. The main challenge in identifying the causal effect of wealth on entrepreneurial entry is finding an exogenous change in household wealth that is unrelated to business opportunities.

In order to overcome the difficulty of estimating the causal effects of household wealth on entrepreneurial decisions, this paper exploits the large and unexpected cash
windfalls gained from big lottery wins. Specifically, we implement a difference-in-differences (DID) design and utilize administrative data covering both lottery winners and business registration in Taiwan. This dataset allows us to track the lottery winning status and business ownership of the same households over time. We compare the trend in business ownership of households receiving large windfalls of more than 1.5 million NT\$ (i.e., around 50,000 US\$) with those only winning less than 15,000 NT\$ (i.e., around 500 US\$), before and after the prize is awarded. Since big lottery wins are almost random events, we show that the treated and untreated families have very similar characteristics and share close trends prior to the receipt of windfalls.



Three key findings emerge. First, our results show that a household's business entry responds positively to a wealth shock. On average, receiving a lottery prize of 1.5 million NT\$ or more can significantly increase a household's probability of starting a new business by 1.5 percentage points (i.e., a 125\% increase from the baseline mean), which is robust to different specifications and various estimated samples.\footnote{One important caveat of our estimates is whether the results based on lottery winners can be applied to the whole population in Taiwan. In order to investigate the external validity of our results, we re-weight the estimated sample to match the characteristics (e.g., age and marital status) of the general population and get similar findings.} The estimated wealth elasticity of starting a business is about 0.20 to 0.36. Moreover, we analyze how a positive wealth shock affects the dynamics of business formation. Our result indicates that the households' entrepreneurial responses show up immediately after winning a big lottery prize, thereby implying that the affected households could already have had a business plan before receiving the cash windfalls but could not implement it due to a lack of funding. The positive effect is both statistically and economically significant at least three years following the receipt of a cash windfall.

Second, in order to understand the possible mechanisms propelling the positive effect of cash windfalls on entrepreneurial entry, we explore which types of households respond the most to a big lottery win.
The subgroup analysis indicates that the impact of a cash windfall on new business creation is more pronounced for households who tend to be liquidity-constrained, such as less wealthy households and those winning lottery prizes during the recession period. We further examine whether the positive wealth effect flattens out after the constraints cease to bind, by breaking down lottery prize amounts into different intervals. Our results suggest that the positive effect of winning a lottery prize on business entry stops increasing when the winning amount surpasses the 30 million NT\$ threshold, which corresponds to the average capital requirement for incorporated companies in Taiwan.\footnote{In Taiwan, the average capital utilized for incorporation was around 32 million NT\$ in 2010.} In addition, the subgroup analysis shows that households establish new businesses in high-capital industries (e.g., transportation industry) or incorporated companies only when receiving a cash windfall of more than 30 million NT\$. These findings suggest that the presence of liquidity constraints could be a crucial barrier to creating businesses that require a certain amount of capital.

Third, we find that winning a lottery prize of 1.5 million NT\$ or more also increases the likelihood of establishing a new firm by 2 percentage points (i.e., a 153\% increase from the baseline mean) for those who already own a business. In contrast, receiving a large cash windfall has a limited impact on the continuation decision of existing firms. Previous literature \citep{robb2014capital, Hadlock,bellon2021personal} has found that liquidity constraints are severe at the time of forming a business but ease over time.\footnote{For instance, \citet{robb2014capital} find that business owners contribute 30\% of the capital raised (approximately 84,900 US\$) in the first year of a new business. However, a proportion of capital raised by owners declines sharply to 14\% (approximately 27,000 US\$) in the fourth year following the start-up. Furthermore, as time passes, owners use more external capital to support firm financing.} Our results are consistent with these findings.

This paper contributes to the existing literature in the following ways. First, we utilize cash windfalls arising from big lottery wins, which provide an unexpected, random, and sizable change in household wealth, to identify the causal effect of household wealth on new business creation.\footnote{\citet{lindh1996self} examined the effect of lottery wins on transitions into self-employment. They found that lottery winnings significantly increase the probability of being self-employed; however, their data was only cross-sectional and did not contain information on the timing of the win, which is crucial for identifying causal effects.} Previous studies have shown that there is a strong positive correlation between wealth and the propensity to start a business, using wealth shocks from receiving an inheritance \citep{evansdavid,evans,holtzdouglas} or local house price appreciation \citep{hurst,Corradin,adelino,schmalz,levine}. However, these events might affect household entrepreneurial activities through other channels besides changing wealth; for example, the receipt of inheritance might also come with a business transfer. Furthermore, house price appreciation not only increases household wealth, but also reflects that local economic conditions are more favorable for starting a business. The nature of the randomized assignment of lottery prizes provides our study with a good exogenous variation in household wealth and a great chance to establish the clean causal effects of wealth on household entrepreneurial decisions.\footnote{The previous literature has studied the effects of lottery wins on labor supply \citep{imbens,furaker2009gambling,cesarini,picchio2018labour}, marriage decision \citep{hankins2011lucky}, individual bankruptcy \citep{hankins}, consumption \citep{kuhn}, stock participation \citep{cesarinistock}, college attendance \citep{bulman2021parental}, and health \citep{apouey2015winning,lindahl2005estimating,kim2021effects,lindqvist2020long}.}

Second, our lottery and business data is drawn from the same set of administrative records, which is almost free from attrition and measurement errors. Several recent studies \citep{bellon2021personal, cespedes2021more, mikhed2019financial}, using wealth shocks from lottery wins, mineral payments, and bonuses of selling jackpot, have collated data manually.\footnote{Instead of using lottery winnings at the individual level, \citet{bermejo2020entrepreneurship} exploits regional variations in cash windfalls generated by the Spanish Christmas Lottery to examine the effect of transitory shocks to local income on firm creation and job creation by startups.} In order to obtain business information, they have to link the lottery or payment data with other datasets, which might inevitably result in some measurement errors or sample attrition. Moreover, our administrative data contains population-wide records on lottery winners and business registration in Taiwan, which has a much larger sample size and a rich set of household characteristics compared to previous studies. 

Third, the windfall amounts in our sample vary greatly in terms of magnitude. This feature allows us to examine the nonlinear effects of wealth on different types of business creation (i.e., high/low industries or incorporated/unincorporated businesses) and help us verify the potential mechanisms behind our findings.  

Finally, this paper also contributes to a small but growing body of literature on serial entrepreneurship \citep{lafontaine2016serial, gompers2010performance, mikhed2019financial, shaw2019productivity, cespedes2021more}. Due to data limitations and the lack of a quasi-experimental design, there is scant causal evidence regarding wealth effects on business owners' entrepreneurial behaviors. Our data containing a vast array of firm registry information enables us to investigate decisions of business entry and exit for serial entrepreneurs, which helps us understand people's entrepreneurial behaviors in a more comprehensive way.

The remainder of this paper is organized as follows. Section \ref{sec:lottery} presents the institutional background to lottery games in Taiwan. Section \ref{sec:data} discusses our data and the estimation sample. Section \ref{sec:empirical} describes our empirical strategy, and in Section \ref{sec:results}, we discuss the main results and examine robustness of our estimates. Section \ref{sec:sub} presents the results of the subgroup analysis, whilst Section \ref{sec:conclusion} provides concluding remarks.	

\section{Institutional Background}
\label{sec:lottery}
In this section, we briefly discuss the institutional details relating to each lottery game in Taiwan. Three lottery games were run by the Taiwanese government during our sample period (i.e., 2004-2018), namely, (1) the Public Welfare Lottery, (2) the Taiwan Receipt Lottery, and (3) the Taiwan Sport Lottery. Our empirical analysis only includes the first two lottery games, since sports lottery winners do not win by ``luck," and any win might be related to professional ability.\footnote{The Taiwan Sports lottery started in 2008 and is the only means of legally betting on sports in the country. There are over ten types of sports and 20 ways to bet, including MLB baseball and NBA basketball from the United States, the major European soccer leagues, Asian baseball, tennis, golf, and the Olympics. According to the games on which one chooses to bet, the odds are different.}

\subsection{Public Welfare Lottery}
The Public Welfare Lottery was initiated by the Ministry of Finance in 1999. According to the “Public Welfare Lottery Issue Act”, the government raises funds for promoting public welfare projects through revenue generated by selling lottery tickets. 

During our sample period, from 2004 to 2018, there were three major types of lottery games: (1) computer-drawn games, (2) scratchcard games, and (3) Keno games. Each type of game has a variety of ways to play. For the computer-drawn games, in general, a player needs to choose a set of numbers, and the goal is to match these to the numbers drawn by the computer. For example, Lotto 6/49 is one of the richest computer-drawn games in Taiwan. In the Online Appendix \ref{app: add_f}, Figure \ref{loto1} displays a Lotto 6/49 card. Players choose six numbers (1-49) at a cost of 50 NT\$ per bet. The jackpot is hit if all six numbers are matched, so the probability of winning the top prize is very low. The jackpot continues to grow until someone wins. 

Figure \ref{scratched} shows a sample scratchcard. These games usually require a player to match numbers or symbols to win the specific prizes. Figure \ref{bingobingo} shows a ticket for a Keno game. The common rule of Keno games is that a player chooses one of ten gameplays and then selects 20 numbers ranging from 1 through 80. The payouts are different depending on the gameplays and the numbers a player chooses. 

\subsection{Taiwan Receipt Lottery}
The Taiwan Receipt Lottery is a bi-monthly, receipt-based invoice lottery, created to encourage legal tax reporting by giving consumers an incentive to purchase at stores that legally report their sales taxes. Figure \ref{receipt} shows eight numbers at the top of the receipt. Every two months, three sets of numbers are randomly drawn by the Ministry of Finance for a special prize, a grand prize, and a first prize. Table \ref{wn} illustrates the prize rules when players' receipt numbers match each kind of prize. The smallest prize is currently 200 NT\$ and the largest 10 million NT\$, but the largest prize before 2011 was 2 million NT\$.

\section{Data and Sample}
\label{sec:data}

\subsection{Data}
Our data were taken from several administrative records: 1) the income statement file, 2) the wealth registry file, 3) the income tax return file, 4) the household registration file, and 5) the business registry file, all of which are provided by Taiwan's Fiscal Information Agency (FIA). All files contain scrambled personal IDs, thereby allowing us to link them at the individual level. 

The data on lottery income is in the income statement file, which reports each payment to an individual on a yearly basis. Some entries based on third-party reported income sources, such as wage income, interest income, pension income, and lottery income. The remaining records contain self-reported information: rental income, business income, and agricultural income. The records on lottery income cover all prize winners winning more than 2,000 NT\$, because only lottery prizes above this amount are taxable and reported to FIA. The income statement file includes the following information: 1) taxpayer ID (i.e., the winner), 2) lottery prize amount, and 3) the bank ID where the prize is redeemed. Since each lottery game has specific banks for prize redemption, we can use bank ID to exclude sport lottery winners. Then, we sum the prizes (public welfare lottery plus receipt lottery) won by individuals on a yearly basis, in order to derive the annual lottery income.

In addition, following \citet{lien2021wealth} and \citet{chu2019variations}, we utilize the income statement file and the wealth registry file to construct individual-level wealth data used for the subgroup analysis. The household registration file provides variables related to demographics, such as gender, year of birth, location of birth, place of residence, year of marriage, and spouse's ID. 

Finally, the records on business ownership are obtained from the business registry file, containing owner ID, firm ID, opening date, location of the firm, and industry. Combining the above files, we are able to understand an individual's business ownership and lottery income in each year.

\subsection{Sample}
\label{sec: sample}
Following \citet{cesarinistock}, the unit of analysis in our empirical specification is a household,
defined as either a lottery winner (i.e., single) or a lottery winner and, if present, his or her spouse (i.e., couple). The treatment group is defined as a household whose annual lottery income is greater than 1.5 million NT\$ (i.e., around 50,000 US\$) in one of the years between 2007 and 2015. Moreover, in order to avoid other wealth shocks affecting a household's entrepreneurial decisions, except for the winning year, the treated households cannot earn more than 15,000 NT\$ (i.e., around 500 US\$)
from a lottery during the study period (i.e., 2004--2018). The control group includes those only winning small prizes, namely, their annual lottery income is less than 15,000 NT\$ in each year. As the control group has never received a big lottery prize, we randomly choose a year in which they received small prizes as a ``placebo" winning year.

To arrive at our estimation sample, we implement the following sample selection process. We first exclude households whose member(s) were dead during the sample period and restrict household members aged between 20 and 60 at the time of the lottery win. Since we focus on people's start-up decisions, the sample is restricted to households who initially (i.e., 3 years before the lottery-winning year) did not own a business. Finally, we follow these households for 7 years (including the year in which they won), starting 3 years before and ending 3 years after winning a big lottery. Note that since the control group has never received a large prize, we randomly assign a ``placebo" winning year to each household in the control group. After applying the above criteria, the number of households belonging to the treatment group and the control group are 2,789 and 1,182,564, respectively.

Table \ref{ds_raw} displays the characteristics of the treatment and control groups before a winning year. Basically, consistent with the random nature of lottery prizes, both groups are quite similar in terms of many characteristics, such as place of residence, household income, employment rate, household wealth, and business ownership in the previous two years. However, we do find that people in the treatment group are slightly older and more likely to get married than those in the control group.
Since these variables could also affect people's start-up decisions, in the robustness check we re-weight the control group to match the distribution of age and marital status in the treatment group.

\section{Empirical Strategy}
\label{sec:empirical}
Our identification strategy is a difference-in-differences (DID) design. This method compares differences in entrepreneurial behaviors between a treatment group and a control group, before and after winning a lottery prize. Since the control group earns very little money from lottery winnings, it is assumed acceptable to remove any shocks other than the big lottery win that might influence the households' decision to start a business. We first use a pre- and a post-DID design by estimating the following regression:

\begin{align}\label{boss_DD}
E_{it}= \alpha Lottery_{i}+\beta Post_{t} + \gamma^{DD}Lottery_{i} \times Post_{t} + \theta_{t} + \nu_{i} + \boldsymbol{X_{it}} \psi  + \varepsilon_{it} 
\end{align}

\noindent where $E_{it}$ represents the outcome of interest: a dummy variable indicating whether a household $i$ owns a business in a given year $t$. $Lottery_{i}$ is a dummy variable indicating that a household $i$ belongs to either the treatment group ($Lottery_{i}=1$) or the control group ($Lottery_{i}=0$). The definition of treatment and control groups has been discussed in the section \ref{sec: sample}. $Post_{t}$ denotes the post-winning period and takes one if a household is observed after a year following a lottery win, and zero otherwise. The key variable in this setting is an interaction term between $Lottery_{i}$ and $Post_{t}$. Thus, $Lottery_{i} \times Post_{t}=1$ means that a household $i$ belongs to the treatment group and is observed after a winning year. The coefficient of interest $\gamma^{DD}$ reflects the average difference in rates of business ownership between households receiving large cash windfalls versus those who do not in the post-winning period, relative to the pre-winning years.

We include the year fixed effect $\theta_{t}$ to control the general trend in entrepreneurial activities and macroeconomic conditions in Taiwan. To make the treatment and control groups more comparable, we also include a set of background characteristics $\boldsymbol{X_{it}}$, some of which are measured at the time prior to winning the lottery: place of residence, employment status, earnings, income, wealth, and lottery prizes won previously. The remaining factors vary according to time, such as age and squared age. The panel structure of the data allows for the inclusion of household fixed effects $\nu_{i}$ to control for any unobservable time-invariant differences between households that may affect their entrepreneurial decisions, such as time preferences (i.e., impatience) or risk attitude. Note that we only control time-varying variables in the specification that includes household fixed effects.\footnote{Therefore, we drop a dummy for treatment group $Lottery_{i}$ and time-invariant covariates when including household fixed effects.} The variable $\varepsilon_{it}$ represents an error term. Finally, since we follow the same households over time, to account for possible serial correlation, the robust standard errors in all regressions are clustered at the household level.

To draw out lottery effects over time, we then extend equation (\ref{boss_DD}) to a dynamic DID model and estimate the following regression:

\begin{align}\label{boss_event}
E_{it}= \alpha Lottery_{i} +\sum_{k\neq-2} \beta_{k} \cdot \mathbf{I}[t-L_{i}=k] +\sum_{k\neq-2} \gamma_{k} \cdot Lottery_{i} \times \mathbf{I}[t-L_{i}=k] + \nonumber   \\ \theta_{t} + \nu_{i} + \boldsymbol{X_{it}} \psi  + \varepsilon_{it} 
\end{align}

\noindent In this specification, all notations are defined in the same way as in equation (\ref{boss_DD}),
except that we replace the post-winning dummy $Post_{t}$ with the time to event dummies $\mathbf{I}[t-L_{i}=k]$. We use $L_{i}$ to represent the lottery winning year. Thus, $\mathbf{I}[t-L_{i}=k]$ is an indicator for being $k$ years from the lottery wins. Note that our dynamic DID model considers a balanced panel of households who we observe every year from 3 years before through 3 years following the time of a big win, and so event time $k$ runs from -3 to +3. Therefore, we use $\mathbf{I}[t-L_{i}=k]$, where $k=-3,-1,0,1,2,3$, to denote event time dummies. For example, $\mathbf{I}[t-L_{i}=1]$ represents a dummy for the first year following a winning year. We omit the event time dummy at $k=-2$ (i.e., the baseline year), suggesting that event time coefficients measure the impacts of cash windfalls relative to the baseline year.

The key variables used for identification in equation (\ref{boss_event}) are a set of event time dummies $\mathbf{I}[t-L_{i}=k]$ interacted with the treatment group dummy $Lottery_{i}$. Coefficients $\gamma_{k}$ represent the difference in the probability of owning a business between the treatment and the control group in the $k^{th}$ year before or after a big win, relative to the difference for the baseline year. We can attribute the distinct trend in business formation between the two groups to the causal effect of receiving a large cash windfall by imposing the identification assumptions below. First, the treatment and control groups' entrepreneurial behavior should follow a common trend in the absence of a windfall. This assumption ensures that our results do not originate from different pre-trends in business entry between the treatment and control groups. We examine the validity of this assumption by using data in the pre-winning period. Second, the composition of the two groups does not change over time. Since the estimation sample is a fixed panel that follows the same households across years, there is no change in group composition during our study period. Under the above assumptions, $\gamma_{k}$ can represent the causal effects of receiving a large cash windfall on business start-ups.

\section{Results}
\label{sec:results}

\subsection{Graphical Evidence}
\begin{figure}[H]
\centering
	\begin{centering}
	\caption{Trend in Business Ownership between Treatment and Control Groups}\label{business_f1}
	\includegraphics[width=0.9\linewidth]{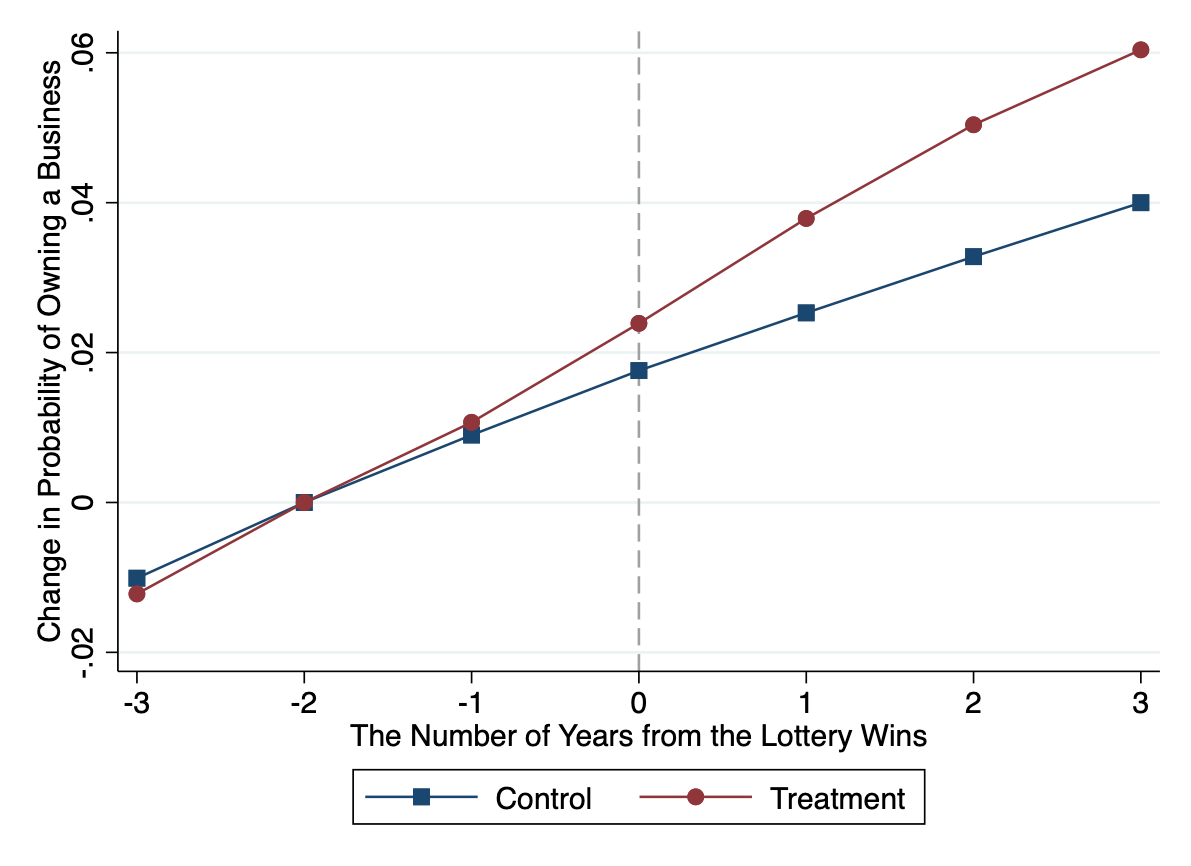} \\
	\end{centering}
	\vspace{0.3cm}
\begin{minipage}{1\linewidth}		
	\fontsize{10}{10pt}\selectfont
	
	\emph{Notes:} This figure displays the change in probability of owning a business from baseline year ($k=-2$) by treatment group (circle symbol)  and control group (square symbol). The sample is restricted to households who initially (i.e., 3 years before the lottery-winning year) did not own a business. The vertical 
axis displays the outcomes at event time $k$ relative to the base year ($k =-2$) for each group. The horizontal axis refers to the number of years from the lottery wins.

\end{minipage}
\end{figure}

Figure \ref{business_f1} displays the evolution of business ownership for the treatment group (i.e., circle symbol) and the control group (i.e., square symbol) from three years before to three years after winning a lottery prize. The vertical axis displays the outcomes at event time $k$ relative to the year of the lottery win ($k = -2$) for each group. Three key insights emerge from this figure. First,
trends in business ownership between the two groups are quite similar prior to the year of winning (i.e., $k=0$). Second, the two series begin to diverge in the winning year, suggesting that treated households start a business right after receiving the large cash windfalls. Third, the increase in business ownership persists at least for three years following a win. Overall, this figure provides clear evidence that the receipt of large cash windfalls does indeed encourage entrepreneurial activities.

\subsection{Average Effects of Cash Windfalls on Entrepreneurial Entry}
Table \ref{main_result} displays estimated coefficients for $Lottery \times Post$ from equation (\ref{boss_DD}). We begin by presenting the estimate from a basic DID regression, controlling for $Lottery \times Post$ (see Column (1)). Then, we gradually control for year fixed effects, average age of household members, pre-winning characteristics, and household fixed effects (see Columns (2) to (5)). The fact that the estimates are quite stable across different specifications is reassuring. All of the estimates are significantly different from zero at the 1\% level. 

Our preferred specification (Column (5)) suggests that the estimated effect of receiving a big lottery prize of more than 1.5 million NT\$ leads to a 1.5 percentage-point increase in the likelihood of owning a business during the three years following this big win. This estimate represents a sizable change (i.e., around a 125\% increase) relative to the baseline probability of 1.2\% for the control group during the pre-winning period. Note that the difference in the average prize for the treatment/control group comparison is about 26.3 million NT\$, which is also very large and corresponds to 491\% of total household wealth for the control group prior to the win.\footnote{According to the information in Table \ref{ds_raw}, the average household wealth of control group is around 3 million NT\$.} Given the above information, our estimate suggests that the wealth elasticity of starting a business is around 0.25. Since prize distribution is highly skewed to the right and contains extreme values that might affect estimated wealth elasticity, and inspired by \citet{olafsson2019borrowing}, \citet{cesarini2016wealth}, and \citet{kumar2009gambles}, we winsorize the amount of lottery prizes at the $98^{th}$ percentile and obtain an estimated wealth elasticity of about 0.36.\footnote{Based on the sample with winsorization, the difference in the average prizes of the two groups become 18.5 million NT\$. This represents a 346\% increase from baseline wealth accumulation.}  Therefore, our estimates suggest that 
households' entrepreneurial decisions are modestly sensitive to a change in wealth.

\subsection{Dynamic Effects of Cash Windfalls on Entrepreneurial Entry}
\begin{figure}[H]
\centering
	\begin{centering}
	\caption{Dynamic DID Estimates}\label{EventStudy}
	\includegraphics[width=0.9\linewidth]{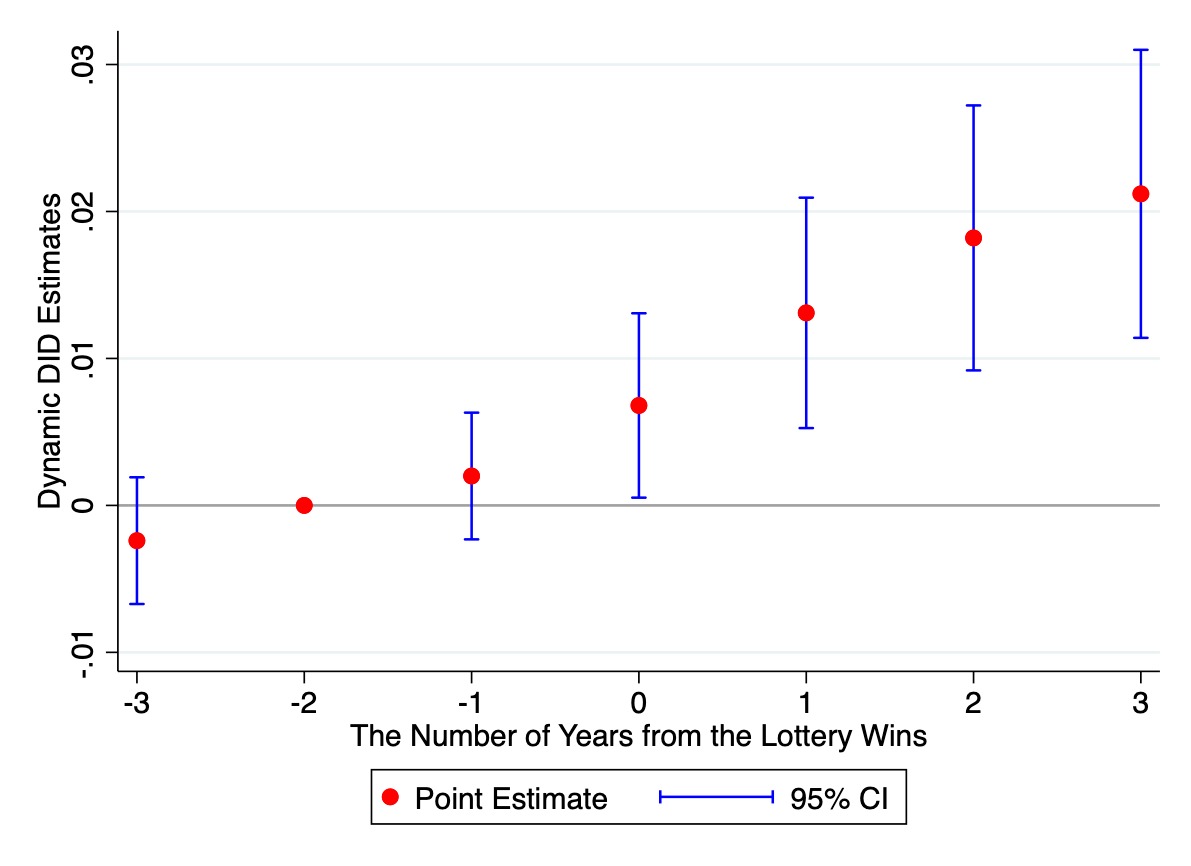} \\
	\end{centering}
	\vspace{0.3cm}
\begin{minipage}{1\linewidth}	
	\fontsize{10}{10pt}\selectfont
	\emph{Notes:} This figure displays the estimated coefficients $\gamma_{k}$ in equation (\ref{boss_event}) from three years before to three years after the time of the lottery wins ($k=-3,-1,0,1,2,3$). The sample is restricted to households who initially (i.e., 3 years before the lottery-winning year) did not own a business. The outcome variable is a dummy variable indicating whether a household $i$ own a business in a given year $t$. The horizontal axis refers to the number of years from the lottery wins. The circle symbol represents the point estimate. The vertical line overlays on the circle symbol is the 95 percent confidence interval.
\end{minipage}
\end{figure}

Figure \ref{EventStudy} displays the results for our dynamic DID model. The vertical axis represents the estimated $\gamma_{k}$ in the equation (\ref{boss_event}), while the horizontal axis denotes the year relative to the time of a big lottery win.

Inspecting this figure provides three key insights. First, the estimated $\gamma_{k}$ at $k=-3,-1$ in the figures are small and statistically insignificant, suggesting that trends in the probability of owning a business run parallel prior to the year of a big win across the two groups. Therefore, the common trend assumption of our DID design is valid. Second, we find that the impact of receiving large cash windfalls on entrepreneurial entry show up immediately in the year after a big win ($k=1$), denoting that the treatment group might already have a business plan at, or before, the time of obtaining significant cash funding but could not implement it due to liquidity constraints. Third, the effects of cash windfalls on business ownership persist and do not disappear in the third year following a big win ($k=3$). In sum, our results indicate that people launch a new business immediately after receiving a large cash windfall and keep the newly established business for at least four years (including the lottery winning year) thereafter.

\subsection{Placebo Tests and Robustness Checks}
\begin{figure}[H]
	\caption{Placebo Tests}	
	\begin{subfigure}{\textwidth}
		\caption{Pre/Post DID Design}\label{Placebo_DD}
		\centering
		\includegraphics[width=0.7\textwidth]{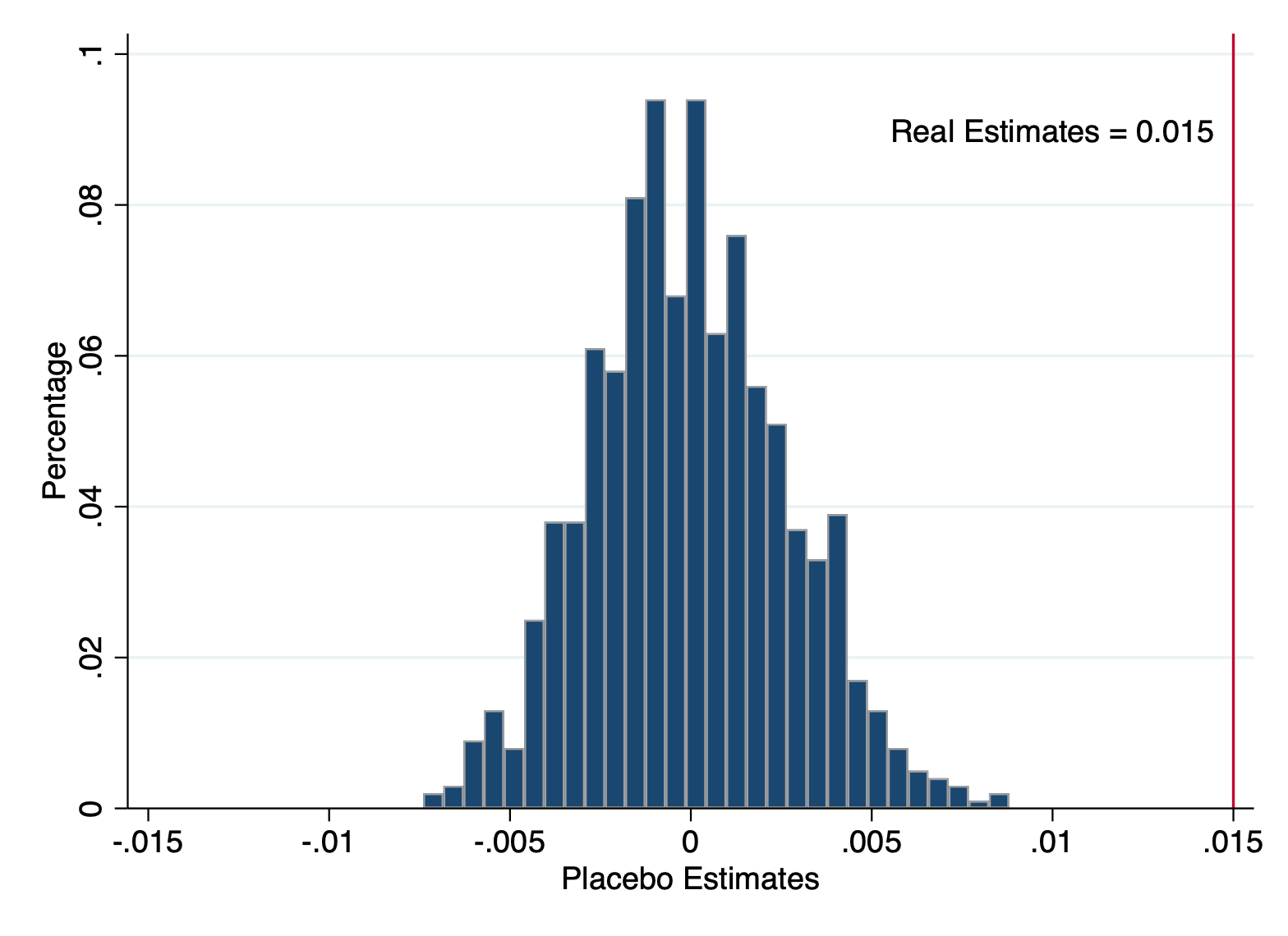}
	\end{subfigure}	
		 \vspace{0.3cm}	
		\begin{subfigure}{\textwidth}	
	 \caption{Dynamic DID Design}\label{Placebo_EventStudy}
  	 \centering
   	 \includegraphics[width=0.7\textwidth]{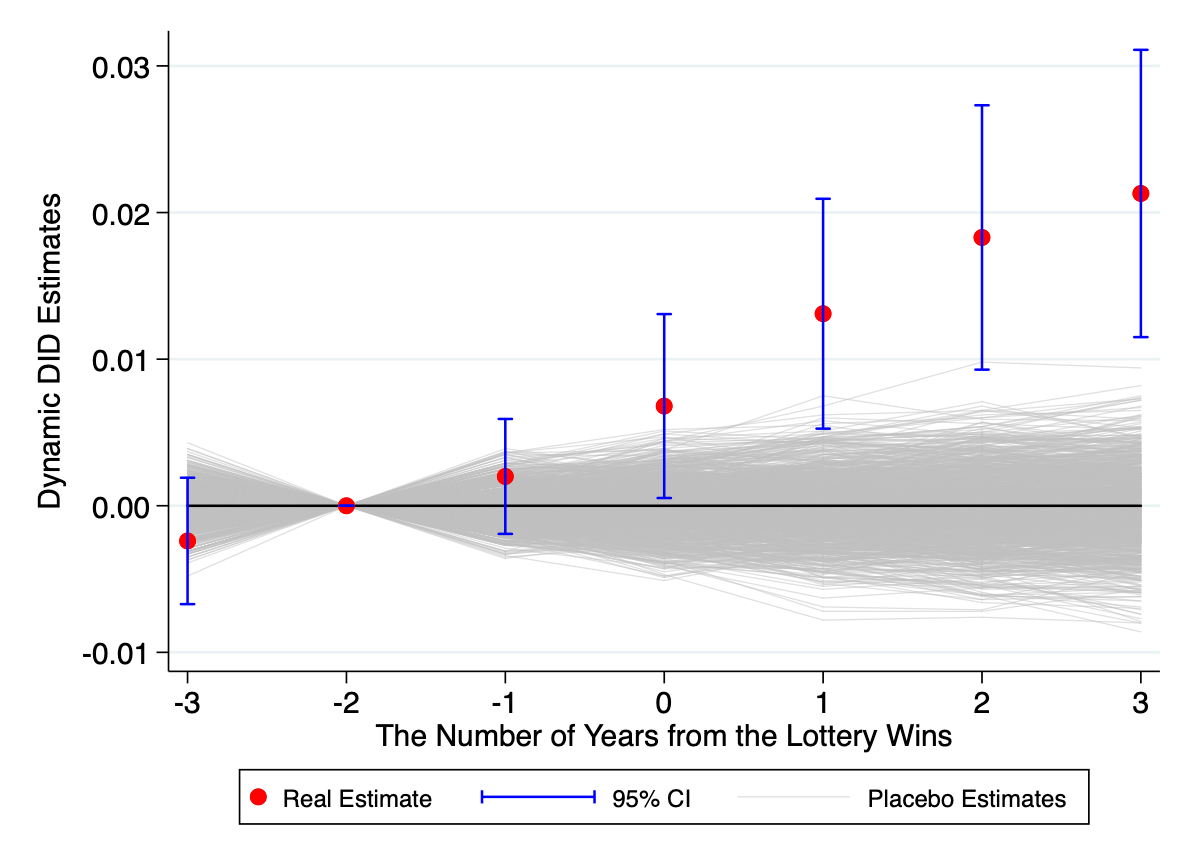}
	\end{subfigure}
		\vspace{0.1cm}
\begin{minipage}{1\linewidth}	
	\fontsize{10}{10pt}\selectfont
	\emph{Notes:}  This figure shows the distribution of 1,000 times placebo estimates. Specifically, we randomly permute lottery prizes and attach them to each household. Then, we use these ``pseudo'' prizes to define treatment/comparison groups and estimate equation (\ref{boss_DD}) and (\ref{boss_event}). We repeat the above procedures 1,000 times to obtain the distribution of pseudo estimates. Figure \ref{Placebo_DD} the real pre/post DID estimate $\gamma_{DD}$ and distribution of pseudo ones (histogram). Figure \ref{Placebo_EventStudy} compares the real estimate (circle symbol) with these fake ones (thin lines with gray color).
\end{minipage}  	
\end{figure}

In this section, we first implement the placebo tests. Specifically, we randomly permute lottery prizes and attach them to each household, following which we use the re-sampling lottery prizes to define the treatment and control groups and the re-estimate equation (\ref{boss_DD}). We repeat the above procedures 1,000 times to derive the distribution of pseudo estimates. Figure~\ref{Placebo_DD} displays the result of this exercise and suggests that the real estimates are much larger than any fake ones. This placebo test indicates that the significant estimates in Table \ref{main_result} should be treated as causal and are not just findings made by chance.

We perform the same placebo test for our dynamic DID analysis by re-estimating equation (\ref{boss_event}) 1,000 times and report estimated $\gamma_{t}$ in Figure~\ref{Placebo_EventStudy}. The red point (circle symbol) represents real estimates, and the gray lines denote 1,000 pseudo ones. The result suggests that the real estimates of $\gamma_{t}$ are much larger than the pseudo ones. Again, the falsification test confirms that significant estimates in the dynamic DID analysis are unlikely to be chance findings. 

Next, we carry out a range of robustness checks for the main results, as illustrated in Table \ref{robust}. In the main specification, we require that the control group can only win a ``small prize", that is, a household earns less than 15 thousand NT\$ from lottery prizes in a given year. Columns (1) and (2) of Table \ref{robust} report the estimate using various ``small prize" definitions---less than 5 thousand NT\$ (Column (1)) or less than 30 thousand NT\$ (Column (2)). We find that our result is robust to this issue.

Our main sample includes the winners of both public welfare lottery and receipt lottery. In Columns (3) and (4) of Table \ref{robust}, we estimate equation (\ref{boss_DD}) separately by type of lottery game. Since the highest prize for the receipt lottery is capped at 10 million NT\$, we find that the entrepreneurial responses to big receipt lottery wins are smaller.

Table \ref{ds_raw} suggests that the sample in the treatment group are older and more likely to be married than those in the control group, which might lead to a distinct trend in entrepreneurial behaviors between the two groups. Following \citet{Schmieder2012}, we use a typical procedure to re-weight the control group to match the distribution of observable characteristics (i.e., average age of households and marital status) in the treatment group.\footnote{We match the following observable characteristics: marital status and 10-year age bins (i.e., dummies indicate age between 20 to 29, 30 to 39, 40 to 49, and 50 to 60)} Table \ref{ds_raw_reweight} in the Online Appendix suggests that after re-weighting, differences in age and marital status between the treatment and comparison groups become smaller and insignificant. The re-weighting estimate is shown in Column (5) and is similar to our main estimate (i.e., 0.015). Furthermore, we conduct one-to-20 matched propensity score matching (PSM) to select similar households in the control group, to match those in the treatment group.\footnote{To be more specific, we first perform a probit regression to predict treatment status, using a set of household characteristics (the same covariates as in equation (\ref{boss_DD})). We then pair each treated household with the 20 control households with the highest propensity scores (i.e., the predicted probability of receiving treatment). Finally, we run the main analysis again, but this time we only use the treated sample and the matched control units.} As expected, Table \ref{ds_raw_psm} reveals that the characteristics of the matched comparison group are very similar to those of the treatment group. Again, we find that our main result is robust to this specification (0.013, See Column (6)).

One important caveat to our analysis is that our sample only consists of people who have won a lottery prize, and so the characteristics of lottery winners could be quite different from those of the general population. Table \ref{descriptive_statistics_sam_pop} in the Online Appendix indicates that the lottery sample was older and more likely to be married than the general population. Besides, the lottery sample also tends to have higher household income and wealth, so, in order to investigate this concern, we re-weight the estimation sample to make their characteristics similar to those of the general population in Taiwan.\footnote{We match the following observable characteristics: marital status, 10-year age bins (i.e., dummies indicate age between 20 to 29, 30 to 39, 40 to 49, and 50 to 60), and household wealth groups (i.e., below zero, equal zero, 1 to 200 thousand NT\$, 200 to 500 thousand NT\$, 500 thousand to 1 million NT\$, 1 to 2 million NT\$, 2 to 5 million NT\$, 5 to 10 million NT\$, 10 to 20 million NT\$, 20 to 50 million NT\$, 50 to 100 million NT\$, 100 to 200 million NT\$, 200 to 500 million NT\$, and greater than 500 million).} After re-weighting, although the differences in observable characteristics between the lottery sample and the population are still statistically significant, due to the large sample size, the magnitudes become much smaller. The magnitudes of these differences, as the proportion of the population mean, are mostly below 10\%. The estimate based on the sample with population re-weighting is shown in Column (7), which is close to our main result (i.e., 0.017).

Recent advancements in DID design \citep{goodman2021difference,de2020two,callaway2021difference,baker2022much,sun2021estimating} have shown that when treatment timing varies, DID estimates might be biased because the control group could contain early-treated units. Our sample consists of nine cohorts who received big lottery prizes in different years (i.e., in a given year during 2007 to 2015). Although this is less of a concern in our research design,\footnote{Our control group are those who have never been treated.} we still adopt a two-step estimation strategy with a bootstrap procedure (CS-DID), proposed by \citet{callaway2021difference}.\footnote{Specifically, the CS-DID estimator is calculated by separately estimating the treatment effect of each cohort (i.e., households winning large lottery prizes in different years), then averaging the results for all possible combinations. The estimator only compares treated households to those who have never been treated and those who have not yet been treated (i.e., the latter-treated cohorts). In addition, this approach uses inverse probability weighting or doubly-robust methods to re-weight the control group, and then it ensures that the observed characteristics of all treated cohorts and the control group are balanced.} Column (8) of Table \ref{robust} shows that the CS-DID estimate is 0.013, which is very close to our main result.

\section{Heterogeneity in Entrepreneurial Responses to Cash Windfalls}\label{sec:sub}

Our findings thus far reflect that, on average, receiving a lottery prize exceeding 1.5 million NT\$ results in an increase in the likelihood of becoming a business owner by 1.5 percentage points (a 125\% increase from the baseline mean). In this section, we investigate the potential mechanisms driving our results by examining several dimensions of heterogeneity in entrepreneurial responses to cash windfalls. 

\subsection{Subgroup Analysis: by Financial Resources and Business Cycle}\label{sec:sub1}
Households could have good business plans (e.g., positive net present value projects) but fail to acquire external funds to start them due to financial friction. According to conventional theories relating to liquidity constraints, a positive wealth shock should have a greater effect on people who are initially more financially constrained \citep{fairlie2012liquidity, bellon2021personal, taylor2001self, johansson2000self, paulson2004entrepreneurship}. Specifically, we use those with less wealth or zero liquid assets prior to a lottery win as the proxy for liquidity-constrained households. Thus, we estimate equation (\ref{boss_DD}) separately by household financial resources. The estimated coefficients of $Lottery \times Post$ are reported in Table \ref{subgroup1}. 

Columns (1)-(2) of Table \ref{subgroup1} display the estimates by household wealth. We define households with less than 1.5 million NT\$ per capita as low-wealth households, and otherwise as high-wealth ones.\footnote{Note that, in order to adjust for household size (i.e., couple or single), we use household income or liquid asset by head rather than the total amount.} The results indicate that the probability of starting a business for low-wealth households significantly increases by 2.3 percentage points when winning a big lottery prize. By contrast, we find that the entrepreneurial responses of high-wealth households are smaller and insignificantly different from zero. A similar pattern can be found in Columns (3)-(4) of Table \ref{subgroup1}, which is based on the level of household liquid assets. 


In the Online Appendix, we also examine heterogeneity in entrepreneurial responses to cash windfalls by various household characteristics. Prior studies have shown that some demographic groups, such as those who are young, single, or unemployed, are more likely to be liquidity-constrained \citep{jappelli1990credit, holtz1993entrepreneurial, xu1995precautionary, benito2006consumption, card2007cash, chetty2008moral, sauer2016rise}. If the presence of financial friction does indeed drive our results, we would expect lottery wins have a greater impact on the entrepreneurial decisions of these constrained people. The estimates in Table \ref{subgroup3} indicate that the positive effect of a lottery win on entrepreneurial behavior appears to be larger for households whose members are below age 45 (Columns (1) and (2)), single (Columns (3) and (4)), and unemployed (Columns (5) and (6)). 

The previous literature \citep{rubinstein2020selection, mikhed2019financial} suggests that people are more likely to obtain external financing in ``boom” rather than in ``bust” periods; thus, liquidity constraints are more relaxed during booms compared with busts.\footnote{A large body of literature has investigated the relationship between entrepreneurship and business cycles \citep{Shleifer1986, CaballeroandHammour1994, BernankeandGertler1989, kiyotaki1997credit, PatrickandHuw2003, Barlevy2007}.} Columns (5) and (6) of Table \ref{subgroup1} examine this prediction by showing
the effect of cash windfalls on business start-ups over business cycles. Similar to other countries around the world, Taiwan experienced a recession in 2008-2009 and then recovered after 2010. Therefore, based on the lottery-winning years, we split the sample into a bust (2007–2009) and a boom (2010–2015) period.\footnote{GDP per capita growth rate is -2.35\% during 2007-2009 (bust period) and 5.07\% during 2010-2015 (boom period), respectively.}

Our results suggest that winning a lottery prize of more than 1.5 million NT\$ can significantly increase the likelihood of becoming a business owner by 3 percentage points in a bust period, which is three times greater than the estimated effect for a boom period. This finding indicates that compared with boom years, liquidity constraints could be more likely to be binding in a bust period. 

Taken together, the above evidence on heterogeneous entrepreneurial responses to cash windfalls is in line with our liquidity constrains interpretation. Households with fewer financial resources, or those who are in the years when fewer external funds are available, respond the most to cash windfalls. 

\subsection{Subgroup Analysis: by Prize Amount and Business Types}\label{sec:sub2}

Some of our results in Table \ref{subgroup1} can also be explained by wealth effects. Since lottery wins substantially increase a household's financial resources, people might become more risk-loving and then start a new business \citep{hurst}. In addition, when households become wealthier, they might consume more non-monetary goods related to business ownership, such as flexible working schedules, the pursuit of dreams, or the power to make business decisions \citep{cespedes2021more,hurst}. In this regard, our findings on stronger entrepreneurial responses to lottery wins for less wealthy households are also in line with our interpretation of the wealth effect.

In this section, we further separate these two mechanisms (i.e., liquidity constraints versus wealth effects) by examining the relationship between cash windfall amount and business type. If liquidity constraints drive the effects of cash windfalls, we might expect the positive relationship between the amount of a lottery win and the probability of owning a business would disappear after the winning prize surpasses the threshold in which the constraints become nonbinding.

In Table \ref{subgroup_prize}, we first investigate whether households' entrepreneurial decisions respond differently according to the lottery prize level. Specifically, we use the same control group but split the treatment group into households winning a lottery prize between 1.5 and 5 million NT\$, between 5 and 30 million NT\$, between 30 and 100 million NT\$, and more than 100 million NT\$. The first two columns of Table \ref{subgroup_prize} indicate that receiving cash windfalls of less than 30 million NT\$ can increase the probability of establishing a business by 1.2 percentage points. Interestingly, we find that the positive impact of winning a big lottery prize on entrepreneurial activities becomes nine times greater when the winning prize exceeds 30 million NT\$. For households whose wins range from 30 to 100 million NT\$, the chances of starting a business increase by 10.2 percentage points (see Column (3)), albeit the estimated effect does not continue to increase and actually becomes smaller (5.4 percentage points, see Column (4)) when the lottery prize surpasses 100 million NT\$. 

Furthermore, we explore why new business creation substantially increases once a lottery prize exceeds 30 million NT\$. It is possible that some business types require a large initial capital investment. For example, in Taiwan, the average capital employed by the transportation industry is around 28 million NT\$ in 2010. If liquidity constraints are a deterrent to starting a business, we speculate that households can establish such high-capital businesses only when they obtain sufficiently large cash windfalls. Inspired by \citet{hurst}, we categorize a business into a high-starting capital industry and a low-starting capital industry.\footnote{High-starting capital industries are the following: mining and quarrying, manufacturing, electricity and gas supply, transportation, communication, finance, insurance, real estate, public utilities, and national defense. The low-starting capital industries are the following: agriculture, forestry, fishing and animal husbandry, construction engineering, wholesale and retail trade, accommodation and food services, education, human health and social work, and arts, entertainment, and recreation} The first two columns in Panel A of Table \ref{subgroup2} suggest that households receiving less than 30 million NT\$ can only create a business in a low-capital industry. Consistent with this result, we find that the receipt of cash windfalls exceeding 30 million NT\$ has larger positive impacts on business creation in high-capital industries (4.9 percentage points) than in low-capital ones (2 percentage points). 

In Columns (3) and (4), we also investigate how the choice of business organization forms interact with the amount of cash windfalls. Previous studies indicate that incorporation requires many more external funds \citep{beck2006small, brixiova2020access, lakuma2019financial}. For instance, in Taiwan, the average capital utilized for incorporation was around 32 million NT\$ in 2010; however, average capital for unincorporated business was only 2 million NT\$. The results in Columns (3) and (4) of Panel A indicate that the receipt of a windfall of less than 30 million NT\$ can only induce the creation of an unincorporated business. In contrast, we find that when households win more than 30 million NT\$, they instead create incorporated businesses. The likelihood of establishing incorporation significantly increases by 4.3 percentage points.

\subsection{Effects of Cash Windfalls on Serial Entrepreneurs' Behaviors}\label{sec:serial}
So far, we have focused on the entrepreneurial responses of households that did not own businesses initially ($k=-3$). The behaviors of serial entrepreneurs have barely been been discussed in previous studies \citep{mikhed2019financial, cespedes2021more}, so in order to provide a more complete picture of how receiving cash windfalls affects people's entrepreneurial decisions, in this section we extend our analysis to households who were already business owners at the beginning of the sample period ($t=-3$). Table \ref{ds_raw_se} in the Online Appendix \ref{app: add_t} displays summary statistics for the treatment and control groups.\footnote{We find that treated households are older, less likely to be couple and living in Taipei, and have more liquid assets than their control counterparts.} 

\begin{figure}[H]
	\caption{Dynamic DID Estimates: Serial Entrepreneurs}\label{EventStudy_serial}
	\begin{subfigure}{\textwidth}
	 \caption{Start a New Business}\label{EventStudy_open}
  	 \centering
   	 \includegraphics[width=0.65\textwidth]{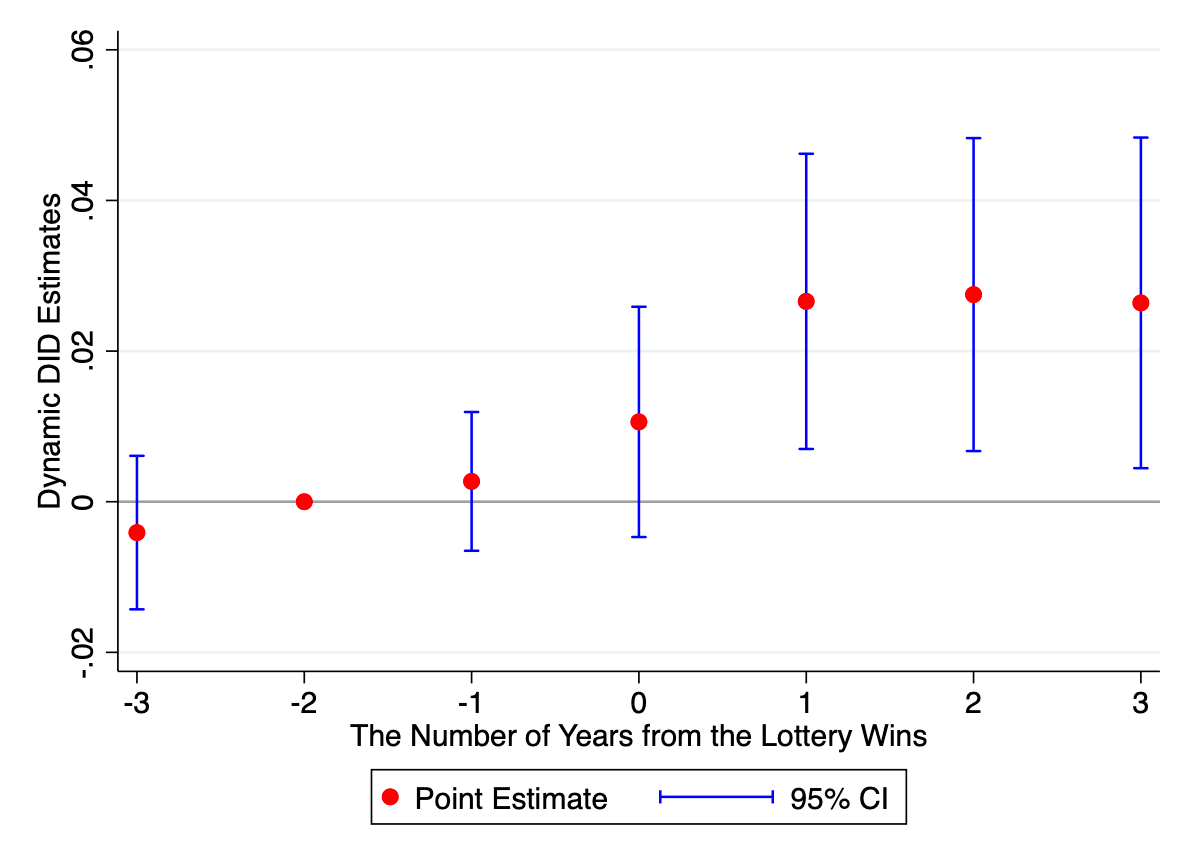}
	 \vspace{0.5cm}
	\end{subfigure}
	\begin{subfigure}{\textwidth}
	\caption{Close the Existing Business}\label{EventStudy_close}
	 \centering
 	\includegraphics[width=0.65\textwidth]{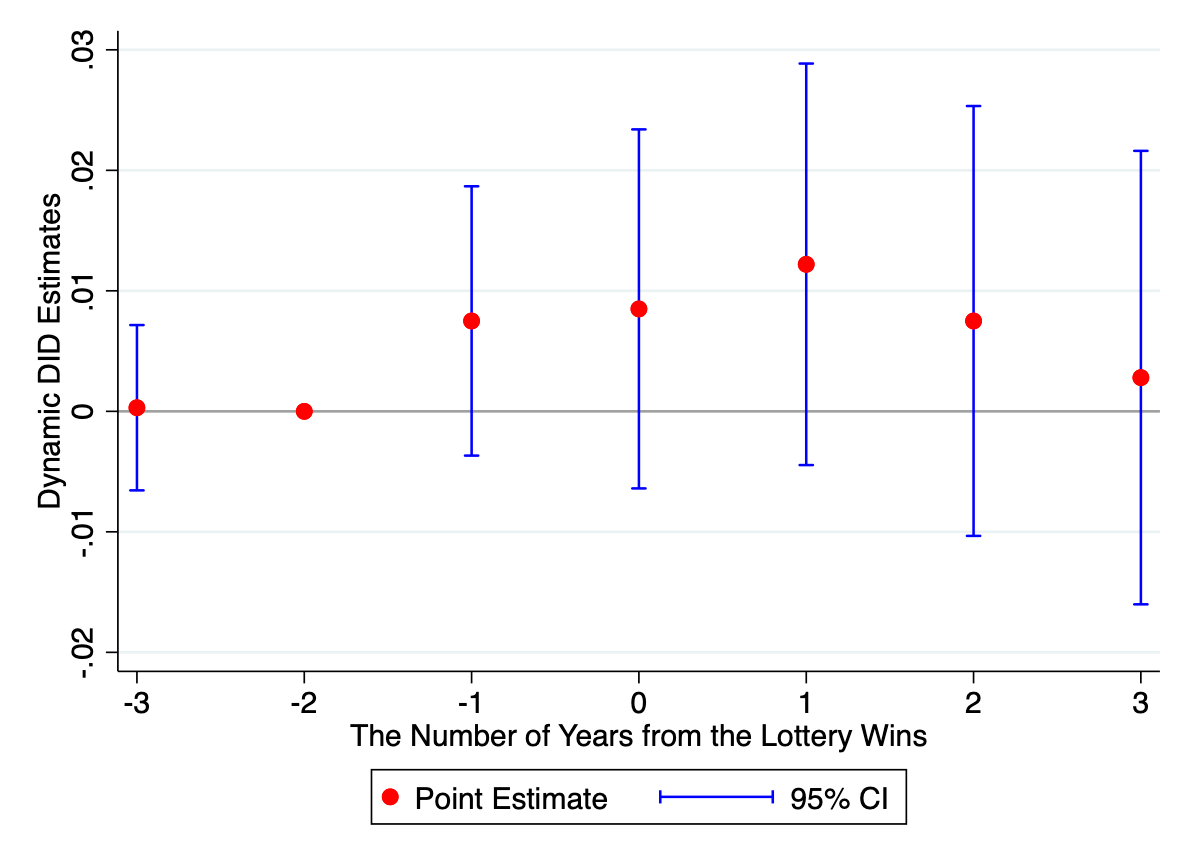}
	\end{subfigure}
	\vspace{0.3cm}
  \begin{minipage}{1\linewidth}	
  	\fontsize{10}{10pt}\selectfont
  	\emph{Notes:} This figure displays the estimated coefficients $\gamma_{k}$ in equation (\ref{boss_event}) from three years before to three years after the time of the lottery wins ($k=-3,-1,0,1,2,3$). The sample is restricted to households who initially (i.e., 3 years before the lottery-winning year) own a business. Outcome in Figure \ref{EventStudy_open} is a dummy variable indicating whether households own businesses established after $k=-3$. Outcome in Figure \ref{EventStudy_close} is a dummy indicating whether or not households close their initial business at $k=-3$. The horizontal axis refers to the number of years from the lottery wins. The circle symbol represents the point estimate. The vertical line overlays on the circle symbol is the 95 percent confidence interval.
  \end{minipage}
\end{figure}

We first investigate whether entrepreneurs start a new business after receiving a large windfall by implementing the same DID regression (i.e., equation (\ref{boss_DD})).\footnote{Note that the outcome is a dummy variable indicating whether households own businesses established after $t=-3$.} Panel A of Table \ref{businessowner} reports the estimated coefficient of $Lottery \times Post$. The estimates are stable across different specifications. Our results indicate that for households already owning a business, the receipt of a cash windfall exceeding 1.5 million NT\$ can significantly increase the likelihood of creating a new business by 2.3 percentage points (see Column(5)). This is a sizable change when considering the baseline mean is only 1.5\%. Figure \ref{EventStudy_open} presents the corresponding dynamic DID estimates based on equation (\ref{boss_event}). We find that entrepreneurs open new businesses immediately after a big lottery win ($k=1$), and the positive effects last for at least three years. 

If entrepreneurs start a new business for expansion or diversification, they might keep their current enterprise open, but it is also possible that they will focus on operating the new business and closing an existing one. Thus, we further examine the effect of cash windfalls on decisions to close existing businesses. We estimate equation (\ref{boss_DD}) by replacing the outcome with a dummy indicating whether or not households close their initial business at $k=-3$. The estimates in Panel B of Table \ref{businessowner} suggest that a big lottery win has insignificant and negligible impacts on closure decisions of current businesses and are robust to different specifications. Similar results can be found in the dynamic DID estimation (see Figure \ref{EventStudy_close}).  

To sum up, our results indicate that receiving a large cash windfall encourages 
entrepreneurs to create new businesses but does not affect their decision to continue with their current enterprise. These findings echo the view that the presence of liquidity constraints is more pronounced at the beginning of forming a business \citep{robb2014capital, Hadlock,bellon2021personal}.

\section{Conclusion}
\label{sec:conclusion}

Utilizing pure wealth shocks induced by lottery wins, this paper examines how the receipt of a cash windfall affects entrepreneurial behavior. We employ a difference-in-differences design by comparing the start-up decisions of households earning more than 1.5 million NT\$ from winning lottery prizes in a given year with those made by people winning less than 15,000 NT\$. Our results show a positive effect of cash windfalls on business entry. The estimated wealth elasticity of starting a business is around 0.25-0.30. Since a lottery win generates a clean resource shock that does 
not confound with other factors, this feature allows us to verify that the positive relationship between wealth and business formation is causal.

We also find that entrepreneurial responses to cash windfalls are more sensitive for households who are likely to be liquidity-constrained. Furthermore, our results indicate that the probability of starting a new business increases as the amount of a lottery win increases, albeit this positive relationship disappears once the winning prizes surpasses the threshold in which the constraints for certain business types are no longer binding. The cutoff is close to the average capital employed for incorporation in Taiwan. Consistent with this result, we find that households start businesses in high-capital industries or incorporate only when they receive a prize greater than 30 million NT\$.

Finally, we investigate the entrepreneurial responses of households initially owning a business. Our results indicate that receiving a large cash windfall encourages 
entrepreneurs to create a new business but does not affect their decision to continue with their current enterprise. Taken together, all the empirical patterns we find are in line with the liquidity constraints model, suggesting that liquidity constraints can be a significant impediment to entrepreneurship.

\newpage
\bibliography{reference}{}	
\bibliographystyle{aea}
\nocite{*}

\newpage
\section*{Tables}

\begin{center}
	\begin{threeparttable}
		\linespread{1}
		\fontsize{8.5}{8.5pt}\selectfont
		\centering\footnotesize
		\caption{Descriptive Statistics for Treatment and Control Group}
		\label{ds_raw}
		\begin{tabular}{@{}lccc@{}}
			\toprule
			& \begin{tabular}[c]{@{}c@{}}Treatment \end{tabular} 
			& \begin{tabular}[c]{@{}c@{}}Control \end{tabular} 
			& \begin{tabular}[c]{@{}c@{}}Difference\\ (Treatment - Control)\end{tabular} \\
			\midrule \midrule
			\textit{\textbf{Household characteristics}}  &  &  &  \\
			~~Average age within household & 39.361  & 36.808 & 2.550*** \\
			& (10.264)	 & (10.551) & {[}0.200{]}\\
			~~Live in Taipei & 0.112 & 0.118 & -0.006 \\
			& (0.315)	& (0.322)	& {[}0.006{]} \\
			~~Couple & 0.505 &	 0.481 & 0.025*** \\
			& (0.500)	& (0.500)	& {[}0.009{]} \\
			~~Employed & 0.777 & 0.787 & -0.010\\
			& (0.416)	& (0.409)	& {[}0.008{]} \ \\
			~~Household earnings (in 1,000 NT\$) & 513.932 & 517.340 & -3.408\\
			& (659.121) & (721.943) & {[}13.662{]}\\
			~~Household income (in 1,000 NT\$) & 601.606 & 606.676 & -5.070\\
			& (772.603) & (872.811) &	 {[}16.516{]} \\
			~~Household wealth (in 1,000 NT\$) & 5,681.412 & 5,352.077 & 329.335\\
			& (12,013.075)	& (18,856.062)	& {[}356.644{]} \\
			~~Household liquidity assets (in 1,000 NT\$) & 1,233.550 & 1,217.813 & 15.738\\
			& (4,163.709)	& (5,041.465)	& {[}95.386{]}  \\
			\midrule
			\textit{\textbf{Lottery variables}} (in 1,000 NT\$) &  &  &  \\
			~~Average amount of lottery prize  & 26,332.226 & 5.161 & 26,327.066*** \\
			& (128,369.608) & (2.294) & {[}117.910{]}  \\
			~~Average amount of lottery prize  &	 18,520.148 & 5.161 &  18,514.986***\\
			~~(Winsorizing at top 2\%)      & (54,872.940) & (2.294) & {[}50.402{]}  \\  \\ 
			~~10 Percentile of lottery prizes &  2,045.827 &	4.043  &  \\
			~~25 Percentile of lottery prizes &  2,208.725&	4.124 &  \\
			~~50 Percentile of lottery prizes &  6,305.833&	4.267 &  \\
			~~75 Percentile of lottery prizes & 8,609.557 &	4.838 &  \\
			~~90 Percentile of lottery prizes & 11,908.185 & 9.315 &  \\
			\midrule
			\textit{\textbf{Outcomes variables}: } &  &  & \\ 
			~~Own a business (1 year before lottery wins) & 0.023 & 0.019 &  0.004 \\ 
			& (0.150) & (0.137) & {[}0.003{]}  \\
			~~Own a business (2 year before lottery wins) & 0.012 & 0.010 &  0.002  \\
			& (0.110) & (0.100) &  {[}0.002{]}  \\  
			\midrule
			\# of households & 2,789 & 1,184,859 &     \\
			\bottomrule
		\end{tabular}
		\begin{tablenotes}[para,flushleft]
		\fontsize{9}{9pt}\selectfont
			Note: The sample is restricted to households who initially (i.e., 3 years before the lottery-winning year) did not own a business. Employed is defined as having positive wage income. Household earnings is calculated as the sum of wage income, business income, and professional income. Household income is calculated as the sum of all income (labor market earnings, interest income, rental income, income from farming, fishing, animal husbandry, forestry, and mining, property transactions income, pension income, and other taxable income), excluding lottery income. Household wealth is calculated as the sum of real estate, capital saving, and stocks less house loan debt. Household liquid asset is defined as the sum of capital savings and stocks. As lottery prize payments are extremely positively skewed, we apply the winsorization method and set extreme outliers equal to 98 percentile. Income, earnings, wealth, and lottery prize amount are adjusted with CPI and displayed in 2016 NT\$ (1 NT\$ $\approx$ 0.033 US\$). Standard deviations in parentheses, and standard errors in brackets.
			*** significant at the 1 percent level,
			** significant at the 5 percent level, and
			* significant at the 10 percent level.
		\end{tablenotes}
	\end{threeparttable}
\end{center}

\newpage
\begin{center}
	\begin{threeparttable}
		\linespread{1.0}
		\fontsize{8.5}{8.5pt}\selectfont
		\centering\footnotesize
		\caption{The Effect of Cash Windfalls on Business Entry}\label{main_result}
		\begin{tabular}{lccccc}
			\toprule
			Dependent Variable: &\multicolumn{5}{c}{Owning a Business} \\
			\cmidrule{2-6}
			& \multicolumn{1}{c}{(1)} &\multicolumn{1}{c}{(2)} & \multicolumn{1}{c}{(3)} 
			& \multicolumn{1}{c}{(4)} & \multicolumn{1}{c}{(5)} \\
			\midrule
			\midrule
			$Lottery \times Post$ & 0.014*** & 0.014*** & 0.016*** & 0.015*** &0.015***  \\
			& (0.003) & (0.003) & (0.003) & (0.003) & (0.004) \\
			\midrule
			Baseline mean & & & 0.012  & & \\
			\# of households & & & 1,187,657 & & \\
			\# of households-years    & & & 8,313,599	 & & \\
			\midrule
			Basic DID controls &  $\surd$ & $\surd$ & $\surd$ & $\surd$ & $\surd$ \\
			Year fixed effect & & $\surd$ & $\surd$ & $\surd$ & $\surd$ \\
			Age effect & & & $\surd$ & $\surd$ & $\surd$ \\
			Pre-winning characteristics & & & & $\surd$ &  \\
			Household fixed effect & & & & & $\surd$ \\
			\bottomrule
		\end{tabular}
		\begin{tablenotes}[para,flushleft]
			\fontsize{9}{9pt}\selectfont
			Note: This table reports coefficients of $Lottery \times Post$ based on equation (\ref{boss_DD}), which stands for the effect of big lottery wins on outcome of interest. The sample is restricted to households who initially (i.e., 3 years before the lottery-winning year) did not own a business.
			The outcome variable is a dummy variable indicating whether a household $i$ own a business in a given year $t$. 			
			The baseline mean is the probability of owning a business for control group in the year right before a lottery win. Column (1) includes a dummy variable indicating a household is belong to the treatment group $Lottery$ and a dummy for the post-winning period $Post$. Column (2) additionally includes calendar year fixed effect. Column (3) additionally includes household average age and its quadratic term.
			Column (4) additionally includes pre-winning characteristics: a set of dummies indicate cities/counties of residence, a dummy indicating couple, a dummy for being employed, total household earnings, total household income, and total household wealth. Note that these covariates are measured in the year right before the lottery-winning year. Column (5) controls for household fixed effects. Standard errors are clustered at the household level and reported in squared brackets.
			*** significant at the 1 percent level,
			** significant at the 5 percent level, and
			* significant at the 10 percent level.
		\end{tablenotes}
	\end{threeparttable}
\end{center}

\newpage
\begin{center}
		\linespread{1.0}
		\begin{threeparttable}
			\fontsize{8.5}{8.5pt}\selectfont
			\centering\footnotesize
			\caption{Robustness Checks}\label{robust}
			\begin{tabular}{@{}lcccccccc@{}}
				\toprule
				Dependent Variable: &\multicolumn{8}{c}{Owning a Business} \\
				\cmidrule(l){2-9}
				& (1) & (2) & (3) & (4) & (5) & (6) & (7)  & (8) \\
				& Cutoff & Cutoff & Welfare & Receipt & Re- & PS  & Population  & CS-DID\\
				& at  5K &  at 30K & Lottery & Lottery &  weighting &  Matching   & Re-  & estimate\\
				&  &  &  &  &   &    & weighting  & \\
				\midrule \midrule
				$Lottery \times Post$ &  0.018*** & 0.017***  & 0.014*** & 0.009*** & 0.015*** & 0.013***  & 0.017*** & 0.013*** \\
				 & {[0.004}{]} & {[}0.003{]} & {[}0.002{]} & {[}0.004{]} & {[}0.002{]} & {[}0.004{]}  & {[}0.002{]}  & {[}0.003{]} \\
				\midrule
				\# of households &  907,957 & 1,377,587  & 489,553 &  690,060 & 1,187,657 & 56,951  &  1,187,657   &  1,187,657 \\
				\# of observations & 6,355,699  &  9,643,109  & 3,,426,871 &  4,830,420 & 8,313,599 & 398,657  & 8,313,599  & 8,313,599\\        
				\bottomrule
			\end{tabular} 
		\begin{tablenotes}[para,flushleft]
				\fontsize{9}{9pt}\selectfont
		Note: This table reports coefficients of $Lottery \times Post$ based on equation (\ref{boss_DD}), which stands for the effect of big lottery wins on outcome of interest. The sample is restricted to households who initially (i.e., 3 years before the lottery-winning year) did not own a business. The outcome variable is a dummy variable indicating whether a household $i$ own a business in a given year $t$. The baseline mean is the probability of owning a business for control group in the year right before a lottery win. All specifications include the same covariates and fixed effects shown in Column (5) of Table \ref{main_result}. Column (1) and (2) reports the estimate using different definition of “small prizes”—less than 5 thousand NT\$ (Column (1)) or less than 30 thousand NT\$ (Column (2)).
			Column (3) only includes people won Welfare Lottery.
			Column (4) only includes people won Receipt Lottery.
		Column (5) reports the estimate based on re-weighting the untreated households to match the distribution of age and marital status of treated ones.
		 Column (6) reports the estimate by utilizing propensity  score  matching (PSM) to construct a comparison  group  with similar pre-winning characteristics   as those in the treatment group.
			Column (7) reports the estimate based on re-weighting the sample to make these characteristics similar to ones of the general population in Taiwan. 
			Column (8) reports the estimate based on a two-step estimation strategy with the bootstrap procedure (CS-DID) proposed by \citet{callaway2021difference}.
			Standard errors are clustered at the household level and reported in parentheses. 
			*** significant at the 1 percent level,
			** significant at the 5 percent level, and
			* significant at the 10 percent level.
		\end{tablenotes}
		\end{threeparttable}
\end{center}

\newpage
\begin{center}
		\linespread{1.0}
		\begin{threeparttable}
			\fontsize{8.5}{8.5pt}\selectfont
			\centering\footnotesize
			\caption{Subgroup Analysis: by Household Wealth and Business Cycle}\label{subgroup1}
			\begin{tabular}{lcccccc}
				\toprule
				Dependent Variable: & \multicolumn{6}{c}{Owning a Business} \\ 
				\cmidrule{2-7}
				& \multicolumn{2}{c}{Wealth} & \multicolumn{2}{c}{Liquidity Assets} &\multicolumn{2}{c}{Business Cycle} \\
				\cmidrule(lr){2-3} \cmidrule(lr){4-5} \cmidrule(lr){6-7}
				& (1) & (2) & (3) & (4) & (5) & (6) \\
				& Below & Above & Below & Above  & Bust      & Boom  \\
				&     1.5M    &    1.5M    &   1.5M     &    1.5M    & 2007-2009 & 2010-2015  \\
				\midrule
				\midrule
				$Lottery \times Post$   & 0.023***& 0.004 & 0.016*** & 0.008 & 0.030*** & 0.010***\\
				&(0.004) &(0.006) &(0.004) &(0.008) &(0.008) &(0.004) \\
				\midrule
				Baseline mean & 0.008 & 0.016& 0.011& 0.016&0.016 & 0.010\\
				\# of individuals &	858,357 &  329,300& 1,047,199& 140,458 & 375,449 & 812,208\\
				\# of observations & 6,008,499& 2,305,100& 7,330,393 & 983,206 &2,628,143 &5,685,456\\
				\bottomrule
			\end{tabular} 
		\begin{tablenotes}[para,flushleft]
			\fontsize{9}{9pt}\selectfont
			Note: This table reports coefficients of $Lottery \times Post$ based on equation (\ref{boss_DD}), which stands for the effect of big lottery wins on outcome of interest. The sample is restricted to households who initially (i.e., 3 years before the lottery-winning year) did not own a business. The outcome variable is a dummy variable indicating whether a household $i$ own a business in a given year $t$. 			
		The baseline mean is the probability of owning a business for control group in the year right before a lottery win. All specifications include the same covariates and fixed effects shown in Column (5) of Table \ref{main_result}.
			Columns (1) and (2) divide household into two groups based on household wealth in the year right before a lottery win. Column (1) includes household whose wealth are below 1.5 million NT\$. Column (2) includes household whose wealth are higher than 1.5 million NT\$. 
			Columns (3) and (4) divide household into two groups based on household liquidity assets. Column (3) includes households whose liquidity assets are 0. Column (4) includes households whose  liquidity assets higher that 0.
			Columns (5) and (6) divide household into two groups based on bust/boom periods. Column (5) includes households who won lottery during 2007 to 2009 (bust period). Column (6) includes households who won lottery during 2010 to 2015 (boom period).
		        Standard errors are clustered at the household level and reported in parentheses.
			*** significant at the 1 percent level,
			** significant at the 5 percent level, and
			* significant at the 10 percent level.
		\end{tablenotes}
		\end{threeparttable}
\end{center}

\newpage
\begin{center}
		\linespread{1.0}
		\begin{threeparttable}
			\fontsize{8.5}{8.5pt}\selectfont
			\centering\footnotesize
			\caption{Subgroup Analysis: By Amount of Prizes}\label{subgroup_prize}
			\begin{tabular}{@{}lcccc@{}}
				\toprule
				Dependent Variable: &\multicolumn{4}{c}{Owning a Business} \\
				\cmidrule(lr){2-5}
				& (1) & (2) & (3) & (4)  \\
				& 1.5M to 5M  & 5M to 30M & 30M to 100M  & Above 100M   \\
				\midrule \midrule
				$Lottery \times Post$ & 0.012** & 0.012** & 0.103** & 0.054**  \\
				& (0.005) & (0.005) & (0.050)& (0.022) \\
				
				\midrule
				\# of households & 1,186,107  &   1,186,227   &  1,184,892 &  1,185,008  \\
				\# of observations & 8,302,749  & 8,303,589  & 8,294,244 & 8,295,056  \\        
				\bottomrule
			\end{tabular}
			\begin{tablenotes}[para,flushleft]
				\fontsize{9}{9pt}\selectfont
				Note: This table reports coefficients of $Lottery \times Post$ based on equation (\ref{boss_DD}), which stands for the effect of big lottery wins on outcome of interest. The sample is restricted to households who initially (i.e., 3 years before the lottery-winning year) did not own a business. The outcome variable is a dummy variable indicating whether a household $i$ own a business in a given year $t$. 			
				The baseline mean is the probability of owning a business for control group in the year right before a lottery win. All specifications include the same covariates and fixed effects shown in Column (5) of Table \ref{main_result}. We use the same control group but split the treatment group into households winning a lottery prize between 1.5 and 5 million NT\$ (Columns (1)), between 5 and 30 million NT\$  (Columns (2)), between 30 and 100 million NT\$  (Columns (3)), and more than 100 million NT\$  (Columns (4)). Standard errors are clustered at the household level and reported in parentheses.
				*** significant at the 1 percent level,
				** significant at the 5 percent level, and
				* significant at the 10 percent level.
		\end{tablenotes}
		\end{threeparttable}
\end{center}

\newpage
\begin{center}
		\linespread{1.0}
		\begin{threeparttable}
			\fontsize{8.5}{8.5pt}\selectfont
			\centering\footnotesize
			\caption{Subgroup Analysis: by Amount of Prize and Business Types}\label{subgroup2}
			\begin{tabular}{@{}lcccc@{}}
				\toprule
				Dependent Variable: & \multicolumn{4}{c}{Owning a Business} \\ 
				\midrule
				& (1) & (2) & (3) & (4)  \\
				& High-Capital & Low-Capital   & Incorporation & Uncorporation  \\
				& Industries   &   Industries  &  &   \\
				\midrule
				\midrule
				\textbf{Panel A:} Below 30M  &  & &  &  \\
				
				$Lottery \times Post$ & 0.003  & 0.008***  &  0.003  & 0.010***  \\
				                      & (0.003) & (0.003) & (0.002) & (0.003)\\
				                      				
				\midrule
				Baseline mean & 0.008 & 0.004 & 0.004 & 0.008 \\
				\# of individuals &\multicolumn{4}{c}{1,187,475} \\
				\# of observations &\multicolumn{4}{c}{8,312,325} \\
				
				\midrule
				\midrule
					\textbf{Panel B:} Above 30M   &  & &  &  \\
					$Lottery \times Post$ & 0.049*** & 0.020*  &  0.043***  & 0.022  \\
				                          & (0.018) & (0.011) & (0.016) & (0.013)\\
				\midrule
				Baseline mean & 0.007 & 0.007 & 0.003 & 0.009 \\
				\# of individuals &\multicolumn{4}{c}{1,185,041} \\
				\# of observations &\multicolumn{4}{c}{8,295,287} \\
				\bottomrule
			\end{tabular} 
			\begin{tablenotes}[para,flushleft]
			\fontsize{9}{9pt}\selectfont
		Note: This table reports coefficients of $Lottery \times Post$ based on equation (\ref{boss_DD}), which stands for the effect of big lottery wins on outcome of interest. The sample is restricted to households who initially (i.e., 3 years before the lottery-winning year) did not own a business. The outcome variable is a dummy variable indicating whether a household $i$ own a specific type of business in a given year $t$. Columns (1) and (2) divide the business types into high and low capital industries. Columns (3) and (4) divide the business into corporation and uncorporation. 			
			The baseline mean is the probability of owning a specific type of business for control group in the year right before a lottery win. All specifications include the same covariates and fixed effects shown in Column (5) of Table \ref{main_result}. We use the same control group but split the treatment group into households winning a lottery prize between 1.5 and 30 million NT\$ (Panel A), and more than 30 million NT\$  (Panel B). Standard errors are clustered at the household level and reported in parentheses.
			*** significant at the 1 percent level,
			** significant at the 5 percent level, and
			* significant at the 10 percent level.
		\end{tablenotes}
		\end{threeparttable}
\end{center}

\newpage
\begin{center}
		\linespread{1.0}
		\begin{threeparttable}
			\fontsize{8.5}{8.5pt}\selectfont
			\centering\footnotesize
			\caption{The Effect of Cash Windfalls on Business Entry and Exit: Serial Entrepreneurs}\label{businessowner}
			\begin{tabular}{@{}lccccc@{}}
				\toprule
				Dependent Variable: & (1) & (2) & (3) & (4)  & (5)\\
				\midrule
				\midrule
				\textbf{Panel A:} Start a New Business  &  & &  &  &\\
				$Lottery \times Post$ & 0.022***  & 0.022***  &   0.022*** & 0.022***  & 0.023**\\
				& (0.008) & (0.008) & (0.008) & (0.008) & (0.008)\\
				Baseline mean & \multicolumn{5}{c}{0.015} \\
				\midrule
				\textbf{Panel B:}  Close the Existing Business   &  & &  &  &\\
					$Lottery \times Post$ & 0.004 & 0.005 &  0.005  & 0.005  & 0.005\\
				& (0.006) & (0.006) & (0.006) & (0.006)  & (0.006)\\
				Baseline mean & \multicolumn{5}{c}{0.056} \\
				\midrule
				\# of individuals & \multicolumn{5}{c}{219,226} \\
				\# of observations & \multicolumn{5}{c}{1,534,582} \\ 			
				\midrule
				Basic DID controls &  $\surd$ & $\surd$ & $\surd$ & $\surd$ & $\surd$ \\
				Year fixed effect & & $\surd$ & $\surd$ & $\surd$ & $\surd$ \\
				Age effect & & & $\surd$ & $\surd$ & $\surd$ \\
				Pre-winning characteristics & & & & $\surd$ &  \\
				Household fixed effect & & & & & $\surd$ \\
				\bottomrule
			\end{tabular} 
		\begin{tablenotes}[para,flushleft]
			\fontsize{9}{9pt}\selectfont
			Note: This table reports coefficients of $Lottery \times Post$ based on equation (\ref{boss_DD}), which stands for the effect of big lottery wins on outcome of interest. The sample is restricted to households who initially (i.e., 3 years before the lottery-winning year, $k=-3$) own a business. Outcome in Panel A is a dummy variable indicating whether households own businesses established after $k=-3$. Outcome in Panel B is a dummy indicating whether or not households close their initial business at $k=-3$. The baseline mean is the average outcome for control group in the year right before a lottery win. Column (1) includes a dummy variable indicating a household is belong to the treatment group $Lottery$ and a dummy for the post-winning period $Post$. Column (2) additionally includes calendar year fixed effect. Column (3) additionally includes household average age and its quadratic term. Column (4) additionally includes pre-winning characteristics: a set of dummies indicate cities/counties of residence, a dummy indicating couple, a dummy for being employed, total household earnings, total household income, and total household wealth. Note that these covariates are measured in the year right before the lottery-winning year. Column (5) controls for household fixed effects. Standard errors are clustered at the household level and reported in squared brackets.
			*** significant at the 1 percent level,
			** significant at the 5 percent level, and
			* significant at the 10 percent level.
					
		\end{tablenotes}
		\end{threeparttable}
\end{center}

\newpage
\section*{Figures}

\newpage

\newpage

\newpage

\newpage

\newpage
\section*{Online Appendix: For Online Publication}

\appendix

\begin{center}
	\setlength{\tabcolsep}{7mm}{
		\Large   
		\begin{tabularx}
			{\linewidth}{l >{\raggedright\arraybackslash}X >{\raggedright\arraybackslash}X}
			Section A & \nameref{app: add_f}  \\
			Section B & \nameref{app: add_t} \\
		\end{tabularx}
	}
\end{center}

\newpage
\setcounter{table}{0}
\renewcommand{\thetable}{A\arabic{table}}
\setcounter{figure}{0}
\renewcommand{\thefigure}{A\arabic{figure}}

\section{Additional Figures}\label{app: add_f}

\begin{figure}[H]
	\centering
		\begin{centering}
			\caption{Computer-Drawn Game - Lott 6/49}\label{loto1}
			\includegraphics[width=0.9\linewidth]{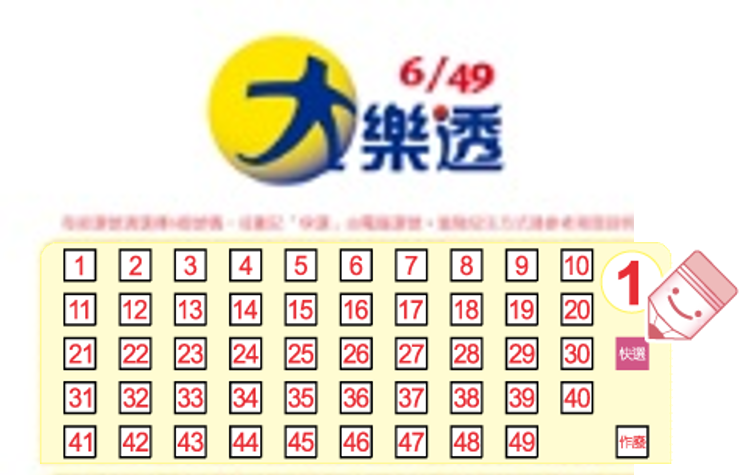} 
	\end{centering}
	\vspace{0.3cm}
  \begin{minipage}{0.9\linewidth}	
	\fontsize{10}{10pt}\selectfont
	\emph{Notes:} This figure displays Lott 6/49 sheet. There exists multiples sections in one sheet. Players choose 6 numbers from section one for one bet. If players want to have more than one bet, they can do the same process in other sections. Players also can choose right column below the section number to let betting machine chooses 6 numbers randomly.
\end{minipage}
\end{figure}

\newpage
\begin{figure}[H]
\centering

	\begin{centering}
		\caption{Scratched Game}\label{scratched}
		\includegraphics[width=0.9\linewidth]{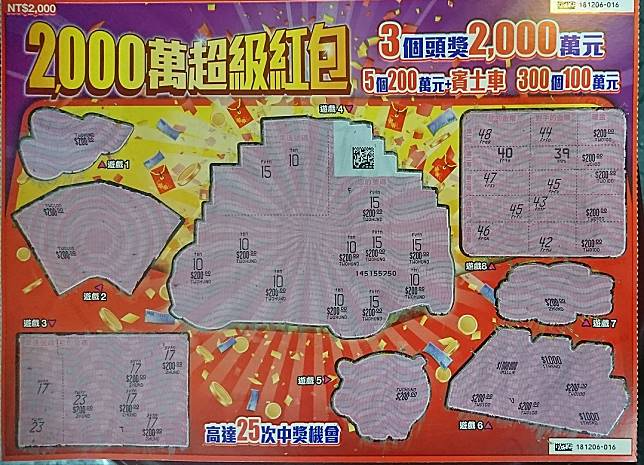}
\end{centering}
	\vspace{0.3cm}
\begin{minipage}{0.9\linewidth}	
	\fontsize{10}{10pt}\selectfont
\emph{Notes:} This figure displays one of the famous scratchcard games. Players can match those numbers or symbols with prizes below to the dash line slots and win the specific prizes. 
\end{minipage}

\end{figure}

\newpage
\begin{figure}[H]
\centering

\begin{centering}
	\caption{Keno Game}\label{bingobingo}
	\includegraphics[width=0.9\linewidth]{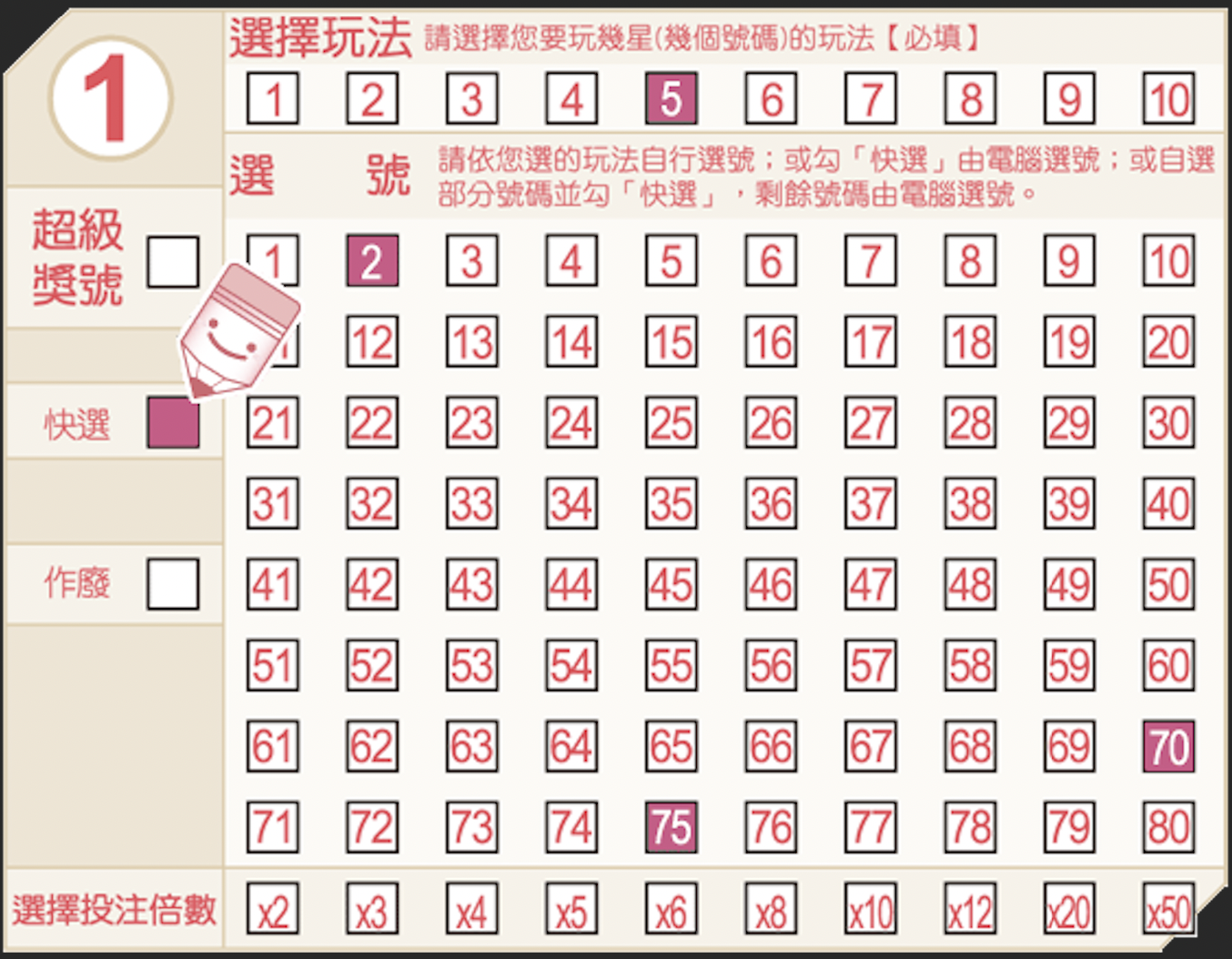} \\
\end{centering}
	\vspace{0.3cm}
\begin{minipage}{0.9\linewidth}	
	\fontsize{10}{10pt}\selectfont
\emph{Notes:} This figure displays keno games. Players first choose one of ten gameplays in first row, then choose 20 numbers from 1 to 80. The prizes will be different according to the gameplays and how many numbers you match.
\end{minipage}

\end{figure}

\newpage
\begin{figure}[H]
\centering

\begin{centering}
	\caption{Taiwan Receipt Lottery}\label{receipt}
	\includegraphics[width=0.5\linewidth]{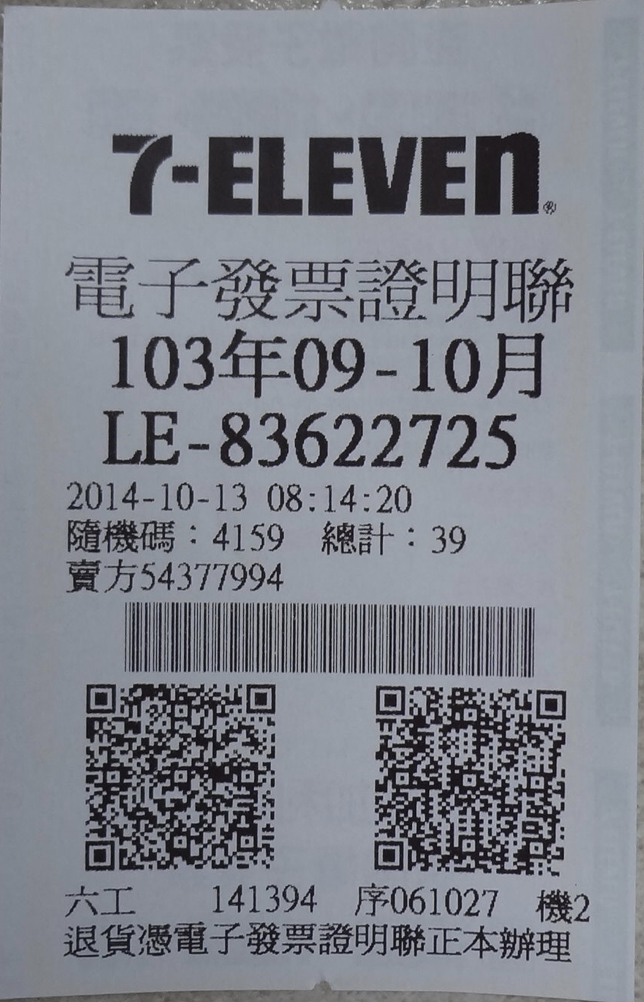}
\end{centering}

\vspace{0.3cm}
\begin{minipage}{0.9\linewidth}	
	\fontsize{10}{10pt}\selectfont
	\emph{Notes:} This figure displays an example of Receipt Lottery, which contains 8 numbers (83622725).
\end{minipage}

\end{figure}

\setcounter{table}{0}
\renewcommand{\thetable}{B\arabic{table}}
\section{Additional Tables}\label{app: add_t}

\begin{table}[H]
	\linespread{1}
	\fontsize{8.5}{8.5pt}\selectfont
	\centering\footnotesize
	\caption{Prizes Rule of Taiwan Receipt Lottery}\label{wn}
	\begin{tabular}{lcl}
		\toprule
		\multicolumn{2}{c}{Prizes (in TWD)}& Matching Winning Number \\
		\midrule
		Special Prize & 10 million & all 8 digits from the special prize number \\
		Grand Prize & 2 million &  all 8 digits from the grand prize number\\
		First Prize  & 200,000  &  all 8 digits from any of the First Prize numbers\\
		Second Prize & 40,000  &  the last 7 digits from any of the First Prize numbers\\
		Third Prize & 10,000  &  the last 6 digits from any of the First Prize numbers\\
		Fourth Prize & 4,000   &  the last 5 digits from any of the First Prize numbers\\
		Fifth Prize & 1,000 &  the last 4 digits from any of the First Prize numbers\\
		Sixth Prize  & 200  &  the last 3 digits from any of the First Prize numbers\\
		Additional Sixth Prize  & 200  &  the last 3 digits from the Additional Sixth Prize number(s)\\
		\bottomrule
	\end{tabular}\\ [0.2cm]
	\begin{minipage}{0.92\textwidth}
		\fontsize{9}{9pt}\selectfont
		Note: This table displays the prizes rule of Taiwan Receipt Lottery. In Figure \ref{receipt}, people can get a Receipt Lottery when they purchased goods, which contains 8 numbers. They match those numbers on the receipt to the ones randomly drawn by Ministry of Finance every two month. 
		
	\end{minipage}
\end{table}%

\newpage
\begin{center}
		\linespread{1.0}
		\begin{threeparttable}
			\fontsize{8.5}{8.5pt}\selectfont
			\centering\footnotesize
			\caption{Subgroup Analysis: by Age, Marital Status, and Employment}\label{subgroup3}
			\begin{tabular}{@{}lcccccc@{}}
				\toprule
				Dependent Variable: & \multicolumn{6}{c}{Owning a Business} \\ 
				\cmidrule{2-7}
				& \multicolumn{2}{c}{\begin{tabular}[c]{@{}c@{}}Age\end{tabular}} 
				& \multicolumn{2}{c}{\begin{tabular}[c]{@{}c@{}}Marital Status\end{tabular}}
				&\multicolumn{2}{c}{\begin{tabular}[c]{@{}c@{}}Employment  \end{tabular}}  \\
				\cmidrule(lr){2-3} \cmidrule(lr){4-5} \cmidrule(lr){6-7} 
				& (1) & (2) & (3) & (4) & (5) & (6)\\
				& Below 45 & Above 45 & Single & Married & Unemployed & Employed \\
				\midrule
				\midrule
				$Lottery \times Post$      &0.020*** &0.005  &0.020***&0.010** &0.022** & 0.014*** \\
				&(0.005) &(0.005)	&(0.005) &(0.005) &(0.010) &(0.004) \\
				\midrule
				Baseline mean & 0.010&	0.015&	0.010&	0.014  & 0.020& 0.013 \\
				\# of individuals &870,112	&317,545  &616,914 &570,743 &143,671&1,043,986 \\
				\# of observations &6,090,784 &2,222,815 &4,318,398 &3,995,201 & 1,005,697&7,307,902 \\
				\bottomrule
			\end{tabular}
	\begin{tablenotes}[para,flushleft]
	\fontsize{9}{9pt}\selectfont
	Note: This table reports coefficients of $Lottery \times Post$ based on equation (\ref{boss_DD}), which stands for the effect of big lottery wins on outcome of interest. The sample is restricted to households who initially (i.e., 3 years before the lottery-winning year) did not own a business. The outcome variable is a dummy variable indicating whether a household $i$ own a business in a given year $t$. 			
	The baseline mean is the probability of owning a business for control group in the year right before a lottery win. All specifications include the same covariates and fixed effects shown in Column (5) of Table \ref{main_result}.
	Columns (1) and (2) divide household into two groups based on average age of household members in the year right before a lottery win. Column (1) includes households whose average age are below 45. Column (2) includes households whose average age are above 45.
	Columns (3) and (4) divide household into two groups based on marital status. Column (3) includes singles. Column (4) includes couples. 
	Columns (5) and (6) divide household into two groups based on employment status. 
	Standard errors are clustered at the household level and reported in parentheses.
	*** significant at the 1 percent level,
	** significant at the 5 percent level, and
	* significant at the 10 percent level.

\end{tablenotes}		
		\end{threeparttable}
\end{center}

\newpage
\begin{center}
	\begin{threeparttable}
		\linespread{1}
		\fontsize{8.5}{8.5pt}\selectfont
		\centering\footnotesize
		\caption{Descriptive Statistics for Treatment and Control Group: After Reweighting}
		\label{ds_raw_reweight}
		\begin{tabular}{@{}lccc@{}}
			\toprule
			& \begin{tabular}[c]{@{}c@{}}Treatment Group \end{tabular} 
			& \begin{tabular}[c]{@{}c@{}}Control Group \end{tabular} 
			& \begin{tabular}[c]{@{}c@{}}Difference\\ (Treatment - Control)\end{tabular} \\
			\midrule \midrule
			\textit{\textbf{Household characteristics}}  &  &  &  \\
			~~Average age within household & 39.361 & 39.122	& 0.240 \\
			& (10.264)	 & (10.483) & {[}0.195{]}\\
			~~Live in Taipei &  0.112 & 0.122 & -0.010* \\
			& (0.315)	& (0.327)	& {[}0.006{]} \\
			~~Couple & 0.505 &	 0.502& 0.003 \\
			& (0.500)	& (0.500)	& {[}0.009{]} \\
			~~Employed & 0.777 & 0.777 & -0.000  \\
			& (0.416)	& (0.416)	& {[}0.008{]} \ \\
			~~Household earnings (in 1,000 NT\$) & 542.802 & 564.941 & -22.139  \\
			& (695.119) & (789.819) & {[}14.723{]}\\
			~~Household income (in 1,000 NT\$) & 601.606 & 629.405 &  -27.799*\\
			& (772.603) & (902.182) &	 {[}16.816{]} \\
			~~Household wealth (in 1,000 NT\$) & 5,681.412 & 6,039.089 &  -357.676 \\
			& (12,013.075)	& (20,341.932)	& {[}378.992{]} \\
			~~Household liquidity assets (in 1,000 NT\$) & 1,233.550 &  1,387.732 & -154.181\\
			& (4,163.709) & (5,432.439)	& {[}101,242{]}  \\
			\midrule
			\textit{\textbf{Lottery variables}} (in 1,000 NT\$) &  &  &  \\
			~~Average amount of lottery prize  &	26,332.226 & 5.161	&  26,327.066*** \\
			& (128,369.608) & (2.300) & {[}117.915{]}  \\
			~~Average amount of lottery prize  &	 18,520.148 & 5.161	& 18,514.987*** \\
			~~(Winsorizing at top 2\%)                & (54,872.940) & (2.300) & {[}50.404{]}  \\ 
			\midrule
			\textit{\textbf{Outcomes variables}: } &  &  & \\ 
			~~Own a business (1 year before lottery wins) & 0.023 & 0.019 & 0.004  \\ 
			& (0.150) & (0.138) & {[}0.003{]}  \\
			~~Own a business (2 year before lottery wins) & 0.012 & 0.010 & 0.002   \\
			& (0.110) & (0.101) &  {[}0.002{]}  \\  
			\midrule
			\# of households & 2,798 & 1,184,859 &     \\
			\bottomrule
		\end{tabular} 
		\begin{tablenotes}[para,flushleft]
		\fontsize{9}{9pt}\selectfont
		Note: The sample is restricted to households who initially (i.e., 3 years before the lottery-winning year) did not own a business. Employed is defined as having positive wage income. Household earnings is calculated as the sum of wage income, business income, and professional income. Household income is calculated as the sum of all income (labor market earnings, interest income, rental income, income from farming, fishing, animal husbandry, forestry, and mining, property transactions income, pension income, and other taxable income), excluding lottery income. Household wealth is calculated as the sum of real estate, capital saving, and stocks less house loan debt. Household liquid asset is defined as the sum of capital savings and stocks. As lottery prize payments are extremely positively skewed, we apply the winsorization method and set extreme outliers equal to 98 percentile. Income, earnings, wealth, and lottery prize amount are adjusted with CPI and displayed in 2016 NT\$ (1 NT\$ $\approx$ 0.033 US\$). Standard deviations in parentheses, and standard errors in brackets.
		*** significant at the 1 percent level,
		** significant at the 5 percent level, and
		* significant at the 10 percent level.
	\end{tablenotes}
	\end{threeparttable}
\end{center}

\begin{center}
	\begin{threeparttable}
		\linespread{1}
		\fontsize{8.5}{8.5pt}\selectfont
		\centering\footnotesize
		\caption{Descriptive Statistics for Treatment and Control Group: PSM}
		\label{ds_raw_psm}
		\begin{tabular}{@{}lccc@{}}
			\toprule
			& \begin{tabular}[c]{@{}c@{}}Treatment Group \end{tabular} 
			& \begin{tabular}[c]{@{}c@{}}Control Group \end{tabular} 
			& \begin{tabular}[c]{@{}c@{}}Difference\\ (Treatment - Control)\end{tabular} \\
			\midrule \midrule
			\textit{\textbf{Household characteristics}}  &  &  &  \\
			~~Average age within household  & 39.361 & 39.318 & 0.044  \\
			& (10.264)	 & (10.117) & {[}0.196{]}\\
			~~Live in Taipei &  0.112 & 0.118 &  -0.006\\
			& (0.315)	& (0.323)	& {[}0.006{]} \\
			~~Couple & 0.505 &	0.512 & -0.007 \\
			& (0.500)	& (0.500)	& {[}0.010{]} \\
			~~Employed & 0.777  & 0.783 & -0.005 \\
			& (0.416)	& (0.413)	& {[}0.008{]} \ \\
			~~Household earnings (in 1,000 NT\$) & 542.802 & 564.943 &  -22.141 \\
			& (695.119) & (745.206) & {[}14.401{]}\\
			~~Household income (in 1,000 NT\$) &  601.606 & 625.575 & -23.969 \\
			& (772.603) & (834.032) &	 {[}16.113{]} \\
			~~Household wealth (in 1,000 NT\$) & 5,681.412 & 5,959.162 & -277.749  \\
			& (12,013.075)	& (16,578.274)	& {[}317.636{]} \\
			~~Household liquidity assets (in 1,000 NT\$) & 1,233.550 & 1,347.688 & -114.137 \\
			& (4,163.709) & (4,528.908) & {[}87.469{]}  \\
			\midrule
			\textit{\textbf{Lottery variables}} (in 1,000 NT\$) &  &  &  \\
			~~Average amount of lottery prize &	26,332.226 & 5.162	&  26,327.064*** \\
			& (128,369.608) & (2.303) & {[}551.545{]}  \\
			~~Average amount of lottery prize  &	 18,520.148 & 5.162	& 18,514.985*** \\
			~~(Winsorizing at top 2\%)                & (54,872.940) & (2.303) & {[235.764}{]}  \\ 
			\midrule
			\textit{\textbf{Outcomes variables}: } &  &  & \\ 
			~~Own a business (1 year before lottery wins) &  0.023 & 0.021 & 0.002  \\ 
			& (0.150) & (0.143) & {[}0.003{]}  \\
			~~Own a business (2 year before lottery wins) & 0.012 & 0.011 & 0.001   \\
			& (0.110) & (0.103) &  {[}0.002{]}  \\  
			\midrule
			\# of households & 2,798 & 54,153 &     \\
			\bottomrule
		\end{tabular} 
		\begin{tablenotes}[para,flushleft]
		\fontsize{9}{9pt}\selectfont
		Note: The sample is restricted to households who initially (i.e., 3 years before the lottery-winning year) did not own a business. Employed is defined as having positive wage income. Total household earnings is calculated as the sum of wage income, business income, and professional income. Total household income is calculated as the sum of all income (labor market earnings, interest income, rental income, income from farming, fishing, animal husbandry, forestry, and mining, property transactions income, pension income, and other taxable income), excluding lottery income. Total household wealth is calculated as the sum of real estate, capital saving, and stocks less house loan debt. Total household liquid asset is defined as the sum of capital savings and stocks. 	As lottery prize payments are extremely positively skewed, we apply the winsorization method and limit extreme values within the 98 percentile. Income, earnings, wealth, and lottery prize amount are adjusted with CPI and displayed in 2016 NT\$ (1 NT\$ $\approx$ 0.033 US\$). Standard deviations in parentheses, and standard errors in brackets.
		*** significant at the 1 percent level,
		** significant at the 5 percent level, and
		* significant at the 10 percent level.
	\end{tablenotes}
	\end{threeparttable}
\end{center}

\begin{center}
	\begin{adjustbox}{angle=90}
		\begin{threeparttable}
			\linespread{1.2}
			\fontsize{9pt}{9pt}\selectfont
			\centering
			\caption{Descriptive Statistics for Estimation Sample and Population}
			\label{descriptive_statistics_sam_pop}
			\begin{tabular}{@{}lcccccccc@{}}
				\toprule
				& \multicolumn{4}{c}{Before Weighting} & \multicolumn{4}{c}{After Weighting} \\
				\cmidrule(r){2-5} \cmidrule(l){6-9}
				& Sample & Population & Difference 	& Proportion(\%)		
				& Sample & Population & Difference & Proportion(\%)\\
				\midrule \midrule
				\textbf{Household characteristics}  &  &  &  & &  \\ 
				~~Average age within household & 37.814 & 36.645 & 1.170*** & 3.19\%&35.765&36.645&-0.880***&-2.40\%\\
				& (10.551) & (11.454) &   & &(11.154)&(11.454)	&& \\
				~~Live in Taipei & 0.121 & 0.113 & 0.008*** & 6.58\%&0.122&	0.113&0.009***	&7.51\% \\
				& (0.326) & (0.317) &   & &	(0.327)	&	(0.317)	&	& \\
				~~Couple & 0.481 & 0.318 & 0.163*** & 51.20\%&	0.318	&	0.318	&	0.000	&	0.00\% \\
				& (0.500) & (0.466) &   & &	(0.466)	&	(0.466)	&		&  \\
				~~Employed & 0.787 & 0.705 & 0.083*** & 11.72\%&	0.759	&	0.705	&	0.054***	&	7.72\% \\
				& (0.409) & (0.456) &   & &	(0.428)	&	(0.456)	&	&  \\	
				~~Household earnings & 548.476 & 415.227 & 133.249*** & 32.09\%&	445.195	&	415.227	&	29.969***	&	7.22\%\\
				~~(in 1,000 NT\$) & (765.759) & (794.375) &   & &	(685.333)	&	(794.375)	&	&  \\
				~~Household income  & 606.664 & 468.087 & 138.577*** &29.61\% &	494.824	&	468.087	&	26.737***	&	5.71\%\\
				~~(in 1,000 NT\$)& (872.588) & (1,167.914) &   &&	(786.407)	&	(1,167.914)	&		&  \\
				~~Household wealth  &5,352.853  & 4,699.304 & 653.549*** & 13.91\%&	4,571.312	&	4,699.304	&	-127.992***	&	-2.72\%\\
				~~(in 1,000 NT\$) & (18,842.864) & (23,670.636) &   &&	(18,405,912)	&	(23,670,636)	&	&	\\
				~~Household liquidity assets  & 1,217.850 & 1,073.532 & 144.318*** & 13.44\%&	1,042.367	&	1,073.532	&	-31.165***	&	-2.90\% \\
				~~(in 1,000 NT\$) & (5,039.575) & (11,130.407) &   & &	(4,734.817)	&	(11,130.407)	&	&	\\
				\midrule
				\# of households & 1,187,654 & 1,187,654 & & & 1,187,654 & 1,187,654 & & \\ 		    
				\bottomrule
			\end{tabular} 
		\begin{tablenotes}[para,flushleft]
		\fontsize{9}{9pt}\selectfont
		Note: The sample is restricted to households who initially (i.e., 3 years before the lottery-winning year) did not own a business. Employed is defined as having positive wage income. Total household earnings is calculated as the sum of wage income, business income, and professional income. Total household income is calculated as the sum of all income (labor market earnings, interest income, rental income, income from farming, fishing, animal husbandry, forestry, and mining, property transactions income, pension income, and other taxable income), excluding lottery income. Total household wealth is calculated as the sum of real estate, capital saving, and stocks less house loan debt. Total household liquid asset is defined as the sum of capital savings and stocks. 	As lottery prize payments are extremely positively skewed, we apply the winsorization method and limit extreme values within the 98 percentile. Income, earnings, wealth, and lottery prize amount are adjusted with CPI and displayed in 2016 NT\$ (1 NT\$ $\approx$ 0.033 US\$). Standard deviations in parentheses, and standard errors in brackets.
		*** significant at the 1 percent level,
		** significant at the 5 percent level, and
		* significant at the 10 percent level.
	\end{tablenotes}
		\end{threeparttable}
	\end{adjustbox}
\end{center}

\begin{center}
	\begin{threeparttable}
		\linespread{1}
		\fontsize{8.5}{8.5pt}\selectfont
		\centering\footnotesize
		\caption{Descriptive Statistics for Treatment and Control Group: Serial Entrepreneurs}
		\label{ds_raw_se}
		\begin{tabular}{@{}lccc@{}}
			\toprule
			& \begin{tabular}[c]{@{}c@{}}Treatment Group \end{tabular} 
			& \begin{tabular}[c]{@{}c@{}}Control Group \end{tabular} 
			& \begin{tabular}[c]{@{}c@{}}Difference\\ (Treatment - Control)\end{tabular} \\
			\midrule \midrule
			\textit{\textbf{Household characteristics}}  &  &  &  \\
			~~Average age within household & 45.334 & 44.456 & 0.878*** \\
			& (8.056)	 & (8.007) & {[}0.306{]}\\
			~~Live in Taipei  &  0.102 & 0.130 &  -0.028**\\
			& (0.302)	& (0.336)	& {[}0.013{]} \\	
			~~Couple & 0.768 & 0.801 & -0.034** \\
			& (0.422)	& (0.399)	& {[}0.015{]} \\
			~~Employed &0.013  & 0.016 & -0.003 \\
			& (0.114)	& (0.126)	& {[}0.005{]} \ \\
			~~Household earnings (in 1,000 NT\$) & 441.878 & 490.669 & -48.791 \\
			& (496.251) & (1,061.617) & {[}40.576{]}\\
			~~Household income  (in 1,000 NT\$) & 679.032 & 659.843 & 19.189 \\
			& (2,462.473) & (1,521.690) & {[}58.379{]} \\
			~~Household wealth  (in 1,000 NT\$) & 13,694.988 &13,110.479 &  584.509 \\
			& (61,939.104)	& (43,888.248)	& {[}1,682.106{]} \\
			~~Household liquidity assets  (in 1,000 NT\$) & 3,416.337 & 2,050.663 & 1,365.674*** \\
			& (45,967.824) & (13,445.317) & {[}523.033{]}  \\
			\midrule
			\textit{\textbf{Lottery variables}}  (in 1,000 NT\$) &  &  &  \\
			~~Average amount of lottery prize  & 22,435.242  & 5.198	& 22,430.043***  \\
			& (118,348.184) & (2.351) & {[}252.976{]}  \\
			~~Average amount of lottery prize  &	 13,851.121 & 	5.198& 13,845.923***\\
			~~(Winsorizing at top 2\%)                & (33,804.376) & (2.351) & {[}72.259{]}  \\ 
			\midrule
			\textit{\textbf{Outcomes variables}: } &  &  & \\ 
			~~Own a business (1 year before lottery wins) & 0.984 &0.980  &   0.004\\ 
			& (0.126) & (0.140) & {[}0.005{]}  \\
			~~Own a business (2 year before lottery wins) & 0.991 & 0.989  &  0.003  \\
			& (0.093) & (0.106) &  {[}0.004{]}  \\  
			\midrule
			\# of households & 685 & 218,541 &     \\
			\bottomrule
		\end{tabular}  			
		\begin{tablenotes}[para,flushleft]
			\fontsize{9}{9pt}\selectfont
			Note: The sample is restricted to households who initially (i.e., 3 years before the lottery-winning year) own a business. Employment is defined as having positive labor market earnings. Household earnings is calculated as the sum of wage income, business income, and professional income. Household income is calculated as the sum of all income (labor market earnings, interest income, rental income, income from farming, fishing, animal husbandry, forestry, and mining, property transactions income, pension income, and other taxable income), excluding lottery income. Household wealth is calculated as the sum of real estate, capital saving, and stocks less house loan debt. Household liquid asset is defined as the sum of capital savings and stocks. As lottery prize payments are extremely positively skewed, we apply the winsorization method and set extreme outliers equal to 98 percentile. Income, earnings, wealth, and lottery prize amount are adjusted with CPI and displayed in 2016 NT\$ (1 NT\$ $\approx$ 0.033 US\$). Standard deviations in parentheses, and standard errors in brackets.
			*** significant at the 1 percent level,
			** significant at the 5 percent level, and
			* significant at the 10 percent level.
		\end{tablenotes}
	\end{threeparttable}
\end{center}

\end{document}